\documentclass{emulateapj}
\newcommand{\myemail}{turrutia@ipac.caltech.edu}

\usepackage{lscape}

\slugcomment{To appear in ApJ, hopefully}
\shorttitle{The FIRST-2MASS Red Quasar Survey II}
\shortauthors{Urrutia et al.}

\begin{document}

\title{The FIRST-2MASS Red Quasar Survey II: An anomalously high fraction of 
LoBALs in searches for dust-reddened quasars\altaffilmark{1}}

\author{Tanya Urrutia\altaffilmark{2,3,4}, 
Robert H. Becker\altaffilmark{3,4},
Richard L. White\altaffilmark{5},
Eilat Glikman\altaffilmark{6},
Mark Lacy \altaffilmark{2},
Jacqueline Hodge\altaffilmark{3},
Michael D. Gregg\altaffilmark{3,4}}

\altaffiltext{1}{Based on observations obtained with the W. M. Keck 
Observatory, which is jointly operated by the California Institute of 
Technology and the University of California}

\altaffiltext{2}{Spitzer Science Center, MS 314-6, California Institute of 
Technology, 1200 E.\ California Boulevard, Pasadena, CA 91125; \myemail
mlacy@ipac.caltech.edu}

\altaffiltext{3}{Department of Physics, University 
of California, One Shields Avenue, Davis, CA 95616; 
hodge@physics.ucdavis.edu }

\altaffiltext{4}{IGPP, L-413, Lawrence Livermore National Laboratory, 
Livermore, CA 94550; bob@igpp.ucllnl.org, gregg@igpp.ucllnl.org}

\altaffiltext{5}{Space Telescope Science Institute, 3700 San Martin Drive, 
Baltimore, MD 21218; rlw@stsci.edu}

\altaffiltext{6}{Astronomy Department, California Institute of Technology, 
Pasadena, CA 91125; eilatg@astro.caltech.edu}

\begin{abstract}
We present results on a survey to find extremely dust-reddened Type-1 
Quasars. Combining the FIRST radio survey, the 2MASS Infrared Survey and the 
Sloan Digital Sky Survey, we have selected a candidate list of 122 potential 
red quasars. With more than 80\% spectroscopically identified objects, well 
over 50\% are classified as dust-reddened Type 1 quasars, whose reddenings 
(E(B-V)) range from approximately 0.1 to 1.5 magnitudes. They lie well off the 
color selection windows usually used to detect quasars and many fall within 
the stellar locus, which would have made it impossible to find these objects 
with traditional color selection techniques. The reddenings found are much 
more consistent with obscuration happening in the host galaxy rather than 
stemming from the dust torus. We find an unusually high fraction of Broad 
Absorption Line (BAL) quasars at high redshift, all but one of them belonging 
to the Low Ionization BAL (LoBAL) class and many also showing absorption 
the metastable FeII line (FeLoBAL). The discovery of further examples of 
dust-reddened LoBAL quasars provides more support for the hypothesis that BAL 
quasars (at least LoBAL quasars) represent an early stage in the lifetime of 
the quasar. The fact that we see such a high fraction of BALs could indicate 
that the quasar is in a young phase in which quasar feedback from the BAL 
winds is suppressing star formation in the host galaxy.
\end{abstract}


\section{Introduction}

\defcitealias{f2m07}{F2M}
\defcitealias{f2m04}{F2M}

Since the identification of 3C 273 \citep{3c273}, our understanding of quasars 
has evolved tremendously. There exists now a relatively robust quasar model in 
which different features of a quasar are explained invoking an 
orientation-model \citep{antonucci,urry}. The orientation model describes the 
quasar by an accretion disk at the center and broad emission line clouds 
directly above and below the disk. The broad-line clouds can be shielded from 
our line of sight by a dusty donut-shaped torus coplanar with the disk. One 
explains whether or not an object is observed to have broad lines in the 
rest-frame optical wavelengths with the viewing angle of the AGN. This model 
has also been successful in explaining the shape of the Extragalactic X-ray 
background spectrum, which peaks at $\sim$ 30keV \citep{treister, gilli}. 
There must be a large population of obscured quasars contributing to it to 
explain its spectrum peaking at such high energies.

The discovery of quasars through color selection techniques based on the 
typically blue spectrum of quasars has been extremely successful. It has 
resulted in a large output of optical spectra such as from the Sloan Digital 
Sky Survey (SDSS, \cite{sc05}), but also in quasar identifications in deep 
multiwavelength surveys (e.g. GOODS \citep{cdfs,goodsir} and NOAODWFS 
\citep{noaodwfs}). 

By 2008 there are well over 100,000 quasars known, with another 100,000 
likely candidates \citep{richards04}. However, the question of completeness of 
the current samples of quasars remains. Recent surveys have shown that 
optically-selected quasars comprise less than half of the total quasar 
population \citep{martinez05,stern,alonso}. Optical quasar surveys tend to 
miss dust-reddened quasars that have been found either with infrared surveys 
\citep{cutri,lacy,f2m04,f2m07}, radio surveys \citep{postman}, hard X-ray 
surveys \citep{polletta} or surveys for high-ionization narrow-line objects 
\citep{zakamska}. Furthermore, the distinction of unobscured / obscured has 
been proven to not be necessarily equivalent to Type 1 / Type 2, as many 
dust-reddened and obscured quasars have been found with broad lines or 
featureless spectra \citep{alonso,martinez06}. Also, due to polarized 
scattering material in the polar directions the Type 1 / Type 2 spectral 
features as an indicator of orientation may be ineffective; quasars with high 
dust-covering factors will be always appear as Type 1 objects optically, 
regardless or orientation \citep{scatter}. In particular, broad absorption 
line quasar (BAL quasar) samples have redder colors than typical quasars 
\citep{sp92,reich} and higher X-ray column densities \citep{gal} also 
suggesting that a large fraction of this population is missing from optically 
selected samples.

This discrepancy in type of obscuration has been observationally recorded, 
especially in objects with different X-ray/mid-IR properties. In some AGN 
which display Type 2 column densities in the X-rays, the mid-infrared dust 
abundances are not correlated with those columns \citep{rigby}. Furthermore, 
there are many high luminosity objects in the mid-infrared with AGN 
signatures, but with a prominent Si absorption feature at 10$\mu$m, that do 
not exhibit strong absorption in the X-rays \citep{hao}.

Current numerical simulations suggest that AGN spend a large phase of their 
lifetime enshrouded by dust from the merging host galaxies 
\citep{hopkins05,li07}. This scenario has already been suggested in earlier 
studies, based on mid- and far-infrared observational evidence 
\citep{sanders}. AGN in a growing phase are thought of as relatively luminous 
and heavily absorbed. Eventually the intense radiation field from the central 
engine destroys the surrounding obscuring material and the AGN shines through 
unabsorbed \citep{dimatteo}. These studies suggest that a non-negligible 
fraction of AGN are obscured by dust in the young merging host galaxy rather 
than a torus. In this picture, the quasar is also responsible for quenching 
some amount of the massive star formation occurring in the bulge of the host 
galaxy. Although there is a strong debate over how much influence this 
``quasar feedback'' has on the host galaxy, these scenarios often invoke 
strong quasar winds that persist well outside the galaxy. BALs, the quasar 
phenomenon that we associate with very strong quasar winds, do not seem to 
have an evolutionary nature, but rather are explained by an ``orientation 
hypothesis'' \citep{weymann,gal07}. However, the special cases of LoBALs (low 
ionization BALs, showing broad absorption troughs in the MgII line) and 
especially FeLoBALS (with absorption troughs in the metastable FeII line) are 
often invoked as classes of BALs that might be explained by a ``youth 
hypothesis'' in which BALs are at an early stage of their evolution in their 
quasar lifetime \citep{voit,bobbal,Trump}. Early studies of LoBALs show them 
to have larger dust covering fractions than standard BAL QSOs and ``normal'' 
QSOs \citep{boroson}. Also for the FeLoBALs, radiative transfer spectral 
synthesis models show them to have large covering fractions for their outflows 
\citep{casebeer} and MIPS photometry shows them to be very luminous in the 
mid-IR, probably due to star formation \citep{farrah}. Unfortunately, so far 
there is not conclusive evidence for PAH emission signifying star formation in 
BAL quasars, but a comprehensive unbiased sample, especially of LoBALs, is 
missing \citep{shi06,shi07}.

Finding dust-obscured quasars in the optical has proven to be a difficult 
task, as dust suppresses mostly UV and optical light. Since we want to study 
quasars at high redshift where the star formation rate and quasar fraction 
peaks, the task is even more difficult, as the optical regime now covers 
mostly rest-frame UV light that is suppressed by several magnitudes from the 
dust. Nevertheless (\cite{f2m04,f2m07}; hereafter F2M) conducted a survey to 
find these dust-reddened and -obscured quasars using the 2MASS near infrared 
survey \citep{2mass} and the FIRST radio survey \citep{first} and following up 
spectroscopically. This F2M survey largely concluded that UVX- and optical 
color-selected sample in the optical miss $\ge$ 10\% of the quasar population 
and that the red quasar population constitutes $>$ 20\% of the total quasar 
population.

In this paper we describe a follow up to the FIRST-2MASS survey described in 
\citetalias{f2m07} to find dust-reddened quasars using the FIRST radio survey 
\citep{first} and the 2MASS Infrared Survey \citep{2mass} but with the 
additional photometry of the SDSS. It expands the original survey area, but 
only explores the objects with the reddest r'-K colors. Throughout this paper 
we use a flat universe with $H_0$ = 70 km$^{-1}$ Mpc$^{-1}$, 
$\Omega_{\Lambda}$ = 0.7 cosmology.

\section{Observations and data analysis}

\subsection{Sample definition}

To create a list of candidate red quasars, the 2MASS All-Sky Point Source 
Catalog (PSC) available at IPAC was matched to the FIRST source catalog within 
2''. In the initial \citetalias{f2m07} selection, the rate of correct matches 
drops significantly beyond 2'', so we adopted this limit as well, reaching 
greater than 95\% completeness at 2''. If there were 2 or more matches within 
2'', then the closest match was taken. This initial matching ensures that we 
have bright infrared sources with radio detections, which will remove most of 
the stars, since they usually do not have strong radio emission. We caution 
that radio selection may be introducing a bias, as quasars with higher radio 
luminosity have a tendency to show redder colors than quasars with very low 
radio luminosities (see Figure 13 of \cite{nano}).

The initial 2MASS catalog contains close to 471 million sources with Near-IR 
magnitudes in the $J$, $H$ and $K_s$ bands. This survey is 99\% complete to a 
magnitude limit at 10$\sigma$ of $J$ = 15.8, $H$ = 15.1 and $K$ = 14.3, 
although fainter sources are included in the catalog all the way down to 
$K \sim 16.0$. The FIRST catalog has more than 810,000 sources with 
$\sim$0.5'' positional accuracy. The 3-minute snapshot integration time 
yields a typical rms of 0.15 mJy, which gives significant detections for the 
catalog at the 5$\sigma$ level of 1 mJy. We obtained a list of 66,699 matches 
between these two catalogs withing 2'' and name the catalog FIRST-2MASS. 

We then matched the FIRST-2MASS catalog to the approximately 215 million 
sources in the DR5 SDSS imaging catalog. We again matched them within 2'' 
using the FIRST position as the comparing position, and if there was more than 
one match within 2'', we chose the closest match. This matching yielded 59,974 
objects; we call the new catalog F2M-SDSS. 

We now filter the F2M-SDSS catalog for objects that are very red, i.e. 
$r'-K>5$. While this selection is different from the \citetalias{f2m04} 
criterion ($R-K>4$), the Sloan $r'$ filter is slightly bluer than the Johnson 
R filter used in the \citetalias{f2m04} color selection and the SDSS uses AB 
magnitudes rather than Vega. Our $r'-K>5$ criterion corresponds to roughly a 
$R-K > 4.5$ for a fairly flat (g'-r')-color as described in \cite{windhorst}. 
Also, we opted to go for even redder objects since we are using the full 2MASS 
source catalog and our resultant FIRST-2MASS catalog is much larger. This 
extremely red color filter leaves us with 603 objects. Since the SDSS 
photometric limits in the r'-band are around 23.1 magnitude, we should with 
good confidence have completeness in the optical, as a color of r'-K $>$ 8 is 
very unlikely. The intrinsic luminosity of objects of objects with such large 
amounts of dust to reach r'-K $>$ 8 colors and still be detected in the 
optical would be very high (e.g. $\sim$ 20 $L^*$ intrinsic luminosity for a 
$z = 1$ red bulge galaxy) and the number density of such luminous objects is 
very small.

\begin{figure}[b!]
\begin{center}
\plotone{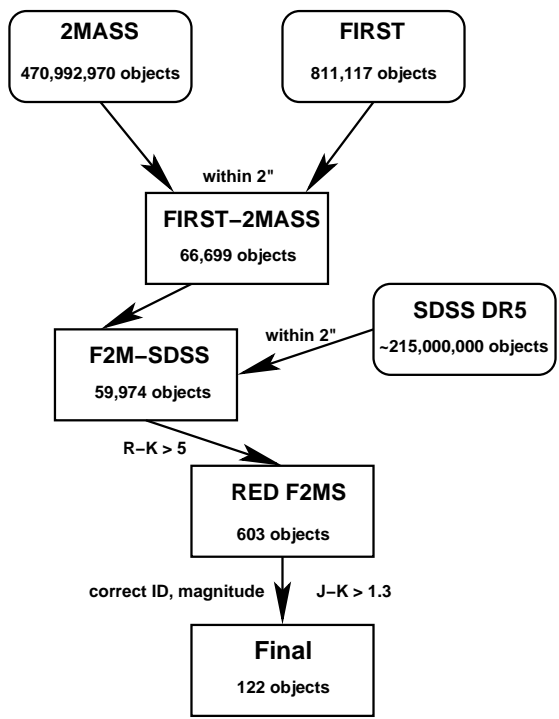}
\caption{Selection process
\label{selection}}
\end{center}
\end{figure} 

\begin{figure*}
\begin{center}
\plotone{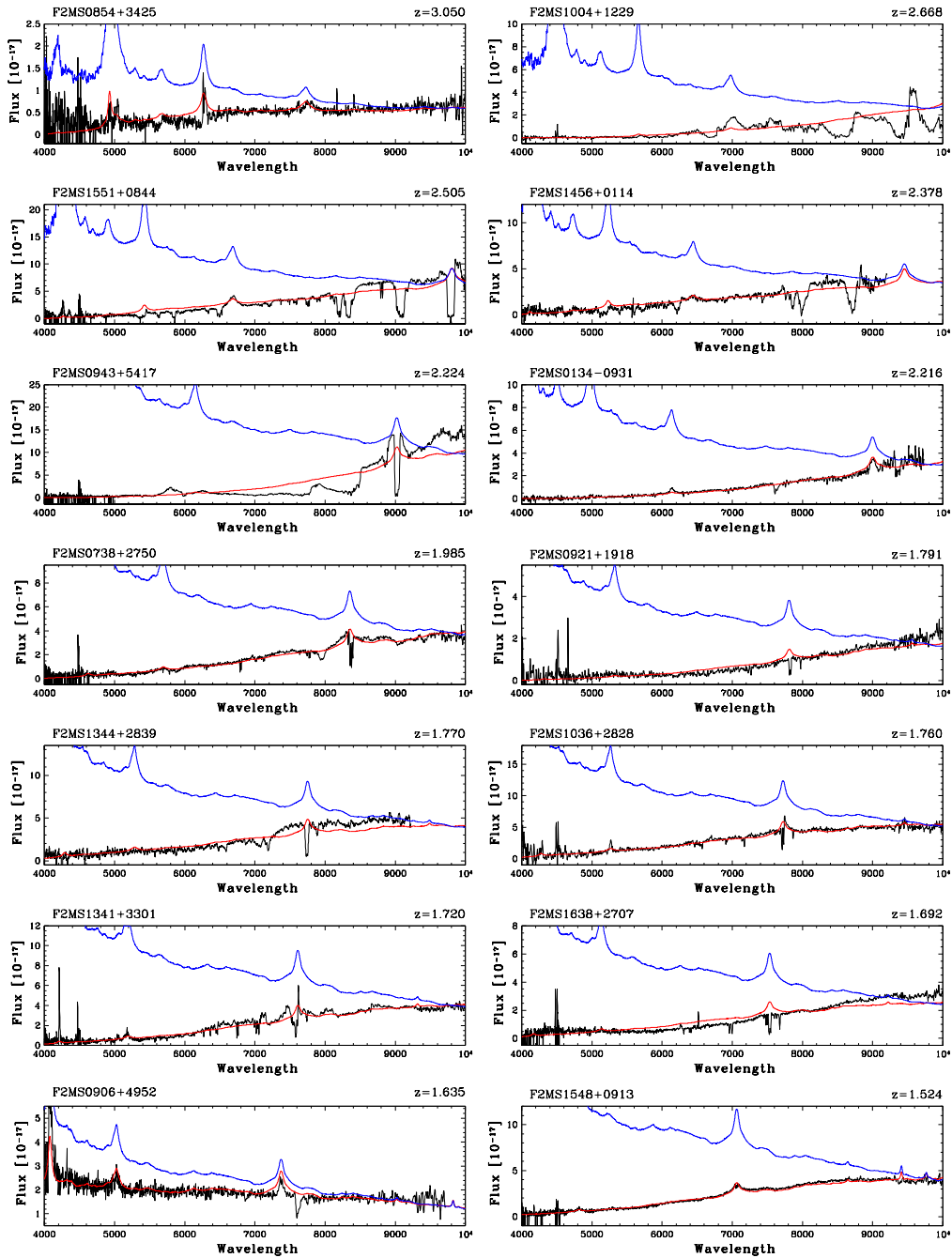}
\end{center}
\caption{Redshift ordered spectra for the red Type 1 QSOs in the F2MS sample. 
The wavelength range shown is from 4,000 to 10,000 \AA . The black line is the 
observed quasar spectrum, the red line is the best fit SMC reddening law 
applied to the FBQS composite. We show the FBQS spectrum as a blue line 
normalized to the reddened composite at 9800 \AA \, for presentation purposes 
only, normally the unreddened spectrum would have much higher flux.
\label{spec}}
\end{figure*}

We inspected these 603 objects by eye. Most of them are bright galaxies, 
which had optical misclassifications or very close double objects in which 
the color filter picked up the faint object but the FIRST coordinate matches 
the bright object much better. The remaining 157 objects are very compact; 
only 2 have Petrosian radii \citep{petrosian} larger than 4''. We decided to 
build in an extra near-IR color filter ($J-K > 1.3$). We have spectra or 
identifications for 18 out of the 35 objects we discarded with this second 
color cut (13 M-Stars, 3 Galaxies and 2 QSOs); this color regime is dominated 
by active radio stars. Quasars only contribute negligibly, so we don't have a 
large bias when discarding objects with this extra color cut. We are, however, 
missing some red quasars in this color range, such as the spectacular example 
of the radio-loud FeLoBAL FBQS1556+3517, which has a $J-K=1.12$ color 
\citep{bob1556}. The $J-K >1.3$ color filter further reduces our catalog to 
122 sources, which is our final F2MS catalog. The whole description is 
presented as a chart in Figure \ref{selection}. The F2MS objects' coordinates 
and photometric properties are presented in Table \ref{observe}.

In summary, the selection criteria are very similar to those determined in 
\citetalias{f2m07}, but we emphasize on only the reddest objects and 
complement with the 5-band color information that SDSS provides. Because of 
the close overlap with the FIRST-2MASS survey, it is only natural that we 
recover many objects already cataloged in \citetalias{f2m07}. Overall, the 
catalogs have 33 (27\%) objects in common.

\setcounter{figure}{1}
\begin{figure*}
\begin{center}
\plotone{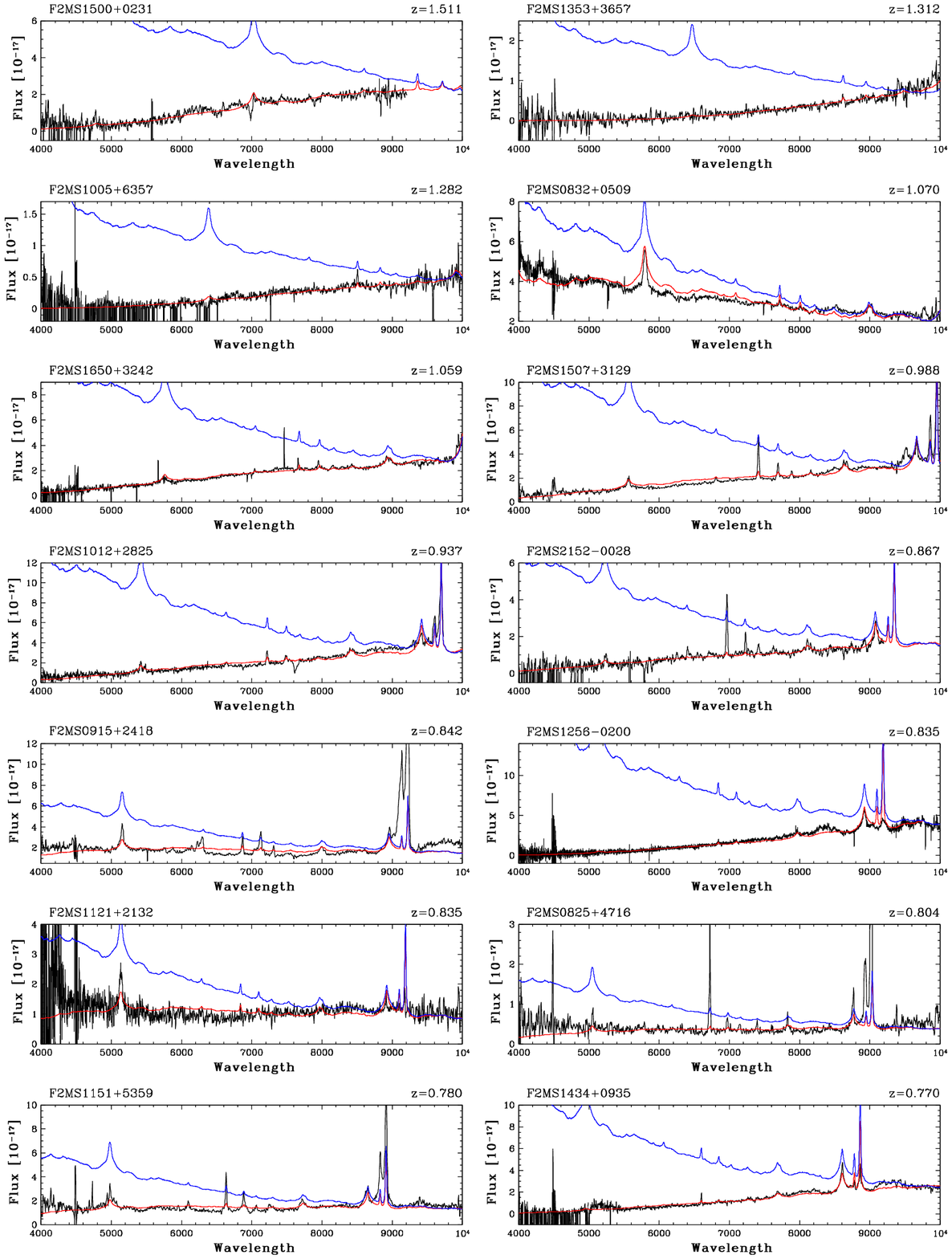}
\end{center}
\caption{cont.}
\end{figure*}

\subsection{Spectral observations}\label{specobs}

To identify the red quasar candidates, we conducted follow-up spectroscopy. 
Twenty of the 122 objects already had spectra taken with the SDSS fibers, so 
we did not duplicate those observations, even though in many cases the signal 
to noise of the SDSS spectra were relatively low. Four objects had already 
been identified spectroscopically through the literature, so we decided 
against repeating those observations, even though we could not analyze the 
respective spectra. As this program has a large overlap with the F2M survey 
and both were done with some temporal overlap, many of the spectra presented 
here are the same as in the F2M survey. Our spectroscopic observations were 
conducted mostly at 10m Keck telescope with the ESI instrument in echellette 
mode with a 1'' slit, giving a spectral resolution of $\sim$1.25 \AA. Five 
objects were observed with the KAST spectrograph on the Shane 3m telescope at 
Lick observatory, at a spectral resolution of $\sim$2.35 \AA. One object was 
observed with the Low Resolution Imaging Spectrometer (LRIS) instrument on the 
Keck 10m telescope and one with the ALFOSC spectrograph on the 2.5m Nordic 
Optical Telescope (NOT). Four of the objects have only Near-IR spectra from 
the SpeX instrument on the IRTF 3m telescope; we adopt their identification, 
but do not analyze their spectra. We now have spectra for exactly 100 objects, 
making spectroscopic identification of the F2MS catalog over 80\% complete.

\setcounter{figure}{1}
\begin{figure*}
\begin{center}
\plotone{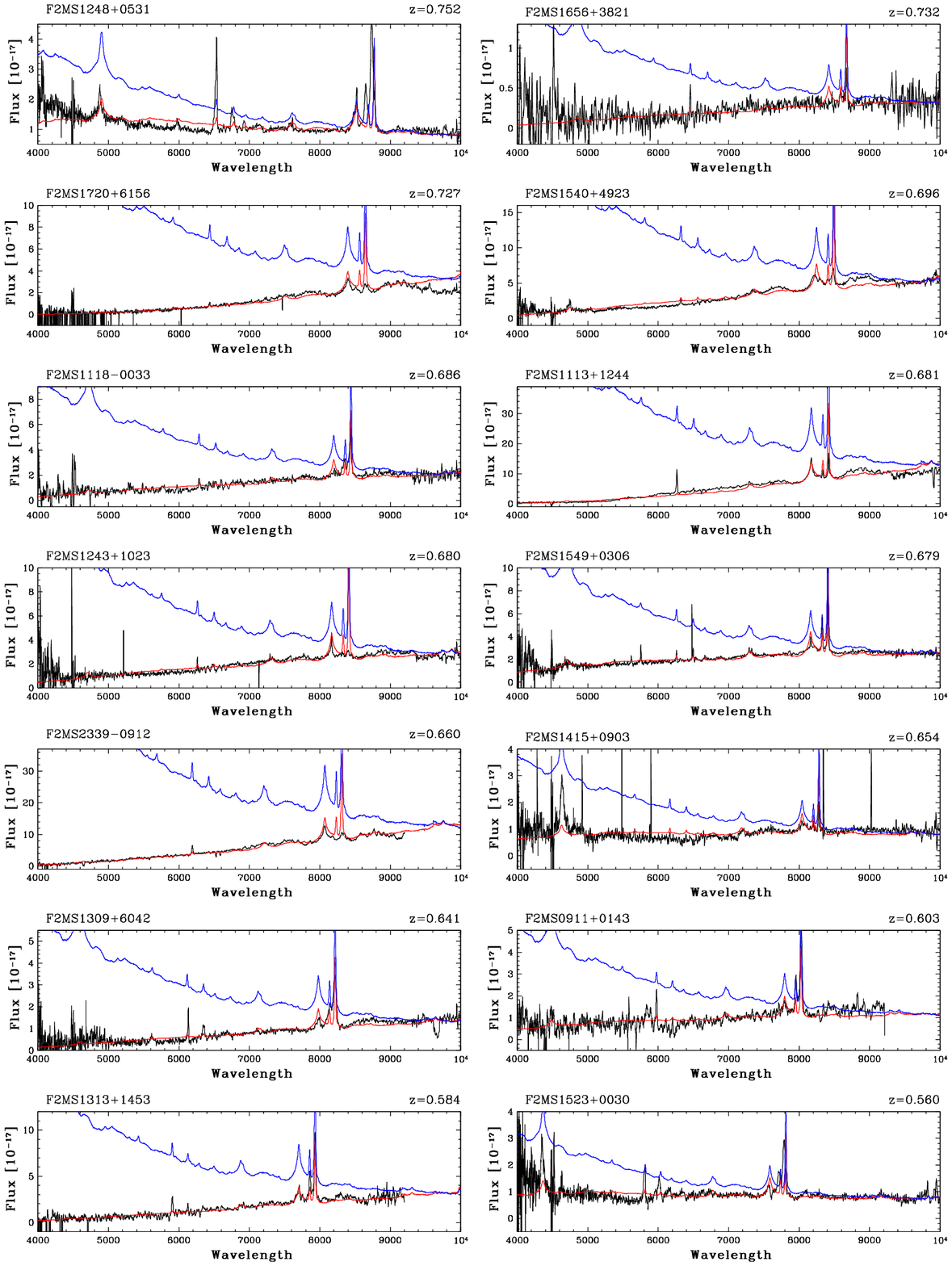}
\end{center}
\caption{cont.}
\end{figure*}

While obtaining the redshift of an object is mostly straightforward, the true 
classification of an object is not trivial. This is especially true for 
spectra that show mixing of various classes of spectra, e.g. a 
post-starburst Balmer complex, but a very strong and broad H$\beta$ line 
indicative of an AGN. Since most of our objects are at moderate redshift, we 
cannot use the ratios around H$\alpha$ to classify the objects with help of 
the BPT diagram. Following the conventions of \citetalias{f2m07}, we 
therefore make the following two cuts to help categorize the spectra: a) If an 
object has an OIII$_{5007}$ / H$\beta$ flux ratio of $>$ 2, it is classified 
as an AGN. b) If the H$\beta$ line of an AGN is broader than 1000 km/s, the 
object is classified as a Broad Line Quasar or Type 1 AGN. Note that we have 
two exceptions to the second rule (F2MS1618$+$3502 and F2MS1656$+$3821), as 
observations in the Near-IR by \citetalias{f2m07} showed broad lines in their 
spectra and these were therefore classified as Type 1 Quasars. For the highest 
redshift objects in which the H$\beta$ line moves out of the optical spectrum, 
we use the MgII line as a proxy for AGN activity. Also at high redshifts we 
have three spectra, which we would not have identified as quasars in the 
optical, but in which the near-IR spectra of \citetalias{f2m07} shows broad 
emission lines and therefore are classified as red quasars (F2M0921$+$2707, 
F2M1313$+$3657 and F2M1344$+$2839). We also did not include known blazars such 
as F2MS1008$+$0621 as QSOs as they were probably included in the F2MS catalog 
due to their high variability rather than because they have an intrinsically 
red color. The identifications, redshifts, comments and references to each 
source are presented in Table \ref{observe} and will be discussed 
further in Section \ref{results}.

\subsection{Reddening}

\setcounter{figure}{1}
\begin{figure*}
\begin{center}
\plotone{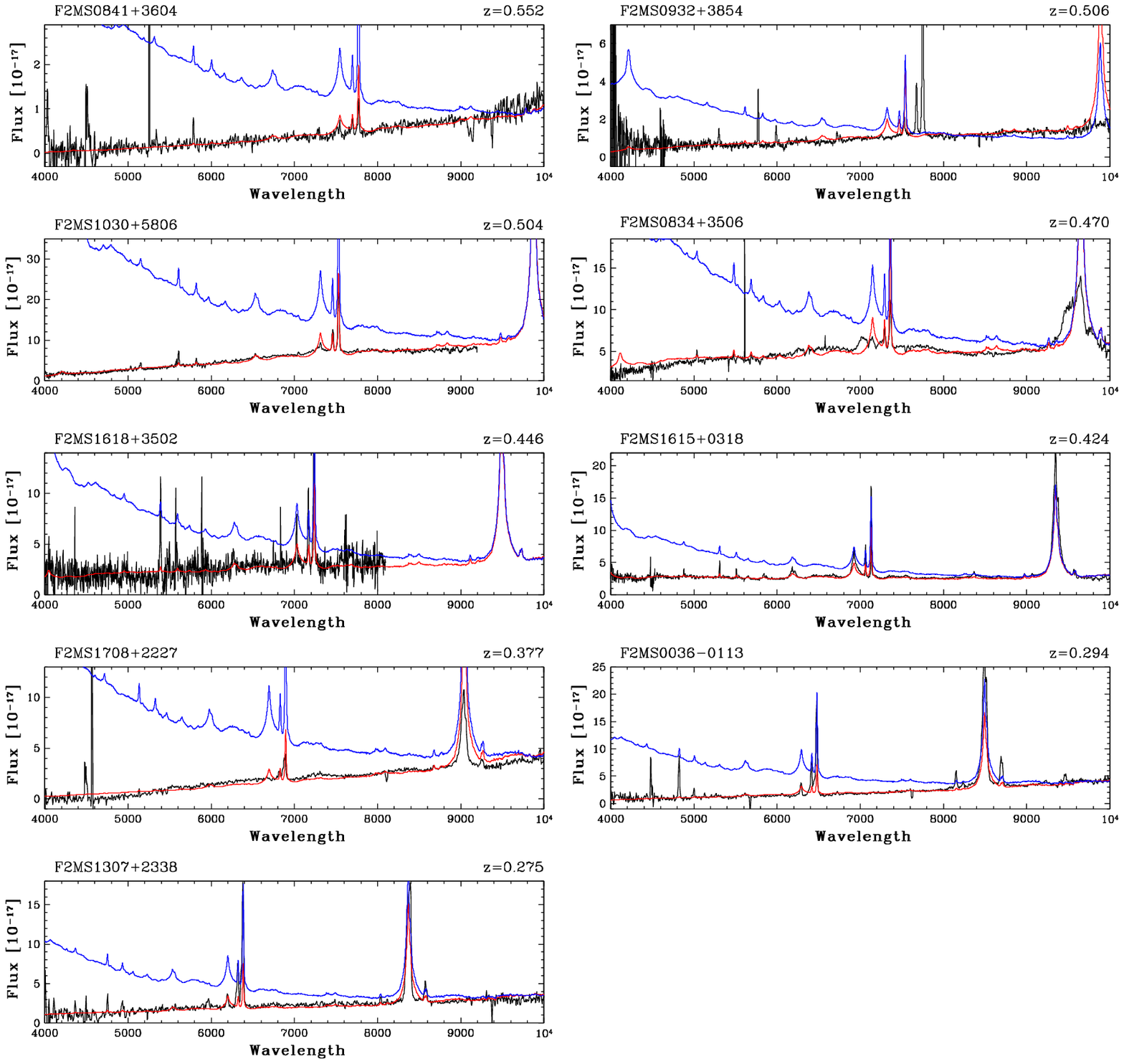}
\end{center}
\caption{cont.}
\end{figure*}

We measure the reddening of the objects classified as Type 1 quasars (QSO in 
Table \ref{observe}) by fitting the red quasar spectra using the FBQS 
composite spectrum \citep{fbqsc} with a SMC dust reddening law. This FBQS 
spectrum was chosen because it represents an unreddened analogue of our 
objects. For the reddening law we use the relation from \cite{fitz99}: 

\begin{equation}
F_o(\lambda) = F_i(\lambda) \, 10^{-E(B-V) \, k(\lambda)} \times C
\end{equation}

\noindent
with $F_i(\lambda)$ the intrinsic, unreddened (FBQS) and $F_o(\lambda)$ the 
observed spectrum and $C$ is a normalization constant. The SMC extinction 
curve $k(\lambda)$ lacks a significant 2175 \AA bump seen in Galactic dust. 
We use the Galactic value of $R_V$ = 3.1, therefore the extinction $A_V$ can 
be deduced by multiplying $E(B-V)$ by $R_V$. Variations in $R_V$ are only 
important above 7000 \AA restframe, so if there are variations of the dust 
grain size, the deduced reddenings will only change for the lowest redshift 
objects. We chose to use the SMC extinction curve for all the quasars to 
deduce their reddenings, as that seems to model extinctions in QSOs the best 
\citep{ri03,postman}. There are some exceptions where the SMC extinction is 
not best \citep{czerny,gaskell}, but for the sake of comparison we apply the 
same dust law to all our Type 1 quasars. For a more detailed discussion on the 
dust law of red quasars and its possible variations, the reader can consult 
Section 5 of Paper I \citetalias{f2m07}. As shown in that discussion, the 
continuum reddening on average was lower than the Balmer reddenings, 
suggesting that the broad line region is even more suppressed than the 
continuum. As the SDSS is designed as an extragalactic survey Galactic 
extinction will not affect the reddening investigation much; Galactic 
extinction values were $<$ 0.12 magnitudes in r' for all objects classified as 
quasars, which will have only a very small effect (E(B-V) $<$ 0.04) on the 
reddening. 

If an object has a reddening $E(B-V) >$ 0.1, we classify it as a red quasar. 
Only two Type 1 quasars are not classified as dust-reddened from our sample 
(F2MS1324$+$0537 and F2MS1102$+$5250), they are most likely highly variable 
objects whose flux varied between the 2MASS and SDSS observation epochs. The 
rest of the Type 1 QSOs had dust-reddenings significant enough to be 
classified as red in the range $E(B-V) = 0.12 - 1.51$ (Table \ref{spectra}). 

These reddenings are significant enough that the very red colors used in the 
selection process could not solely have come from a ``red'' spectral slope as 
is the case in some of the SDSS red quasars \citep{ri03}. As discussed in 
\citep{f2m04,f2m07} the reddening is also unlikely to come from a strong 
synchrotron component, so overall the red spectral slope is consistent with 
dust-reddening. In addition, the reddenings can even be underestimated in some 
cases as the quasar is so obscured in the UV that there is significant host 
galaxy contribution to the spectrum \citep{redqso-hst}. The spectra of the red 
quasars can be seen ordered by decreasing redshift in Figure \ref{spec}.

\section{Results}\label{results}

\begin{deluxetable*}{lcccc}
\tabletypesize\footnotesize
\tablecaption{Spectral properties of the Quasars \label{spectra}}
\tablehead{
\colhead{Source} & \colhead{Redshift} & \colhead{E(B$-$V)} & \colhead{H$\beta$ width} & 
\colhead{Absolute Mag.} \\
\colhead{} & \colhead{\it{z}} & \colhead{(rest)} & \colhead{(km s$^{-1}$)} & \colhead{$M_i$}
}
\startdata
F2MS0036$-$0113 & 0.294 & 0.96 $\pm$ 0.10 & 1150 $\pm$ 100 & $-$23.19 \\
F2MS0134$-$0931 & 2.216 & 0.84 $\pm$ 0.11 & ...            & $-$31.44 \\
F2MS0738$+$2750 & 1.985 & 0.67 $\pm$ 0.11 & ...            & $-$29.76 \\
F2MS0825$+$4716 & 0.804 & 0.52 $\pm$ 0.10 & 1225 $\pm$ 150 & $-$27.65 \\
F2MS0832$+$0509 & 1.070 & 0.14 $\pm$ 0.04 & ...            & $-$23.71 \\
F2MS0834$+$3506 & 0.470 & 0.58 $\pm$ 0.05 & 2350 $\pm$ 100 & $-$24.25 \\
F2MS0841$+$3604 & 0.552 & 1.34 $\pm$ 0.11 & 1500 $\pm$ 500 & $-$27.16 \\
F2MS0854$+$3425 & 3.050 & 0.20 $\pm$ 0.17 & ...            & $-$27.23 \\
F2MS0906$+$4952 & 1.635 & 0.12 $\pm$ 0.07 & ...            & $-$24.38 \\
F2MS0911$+$0143 & 0.603 & 0.63 $\pm$ 0.17 & 1650 $\pm$ 300 & $-$24.20 \\
F2MS0915$+$2418 & 0.842 & 0.36 $\pm$ 0.12 & 1900 $\pm$ 200 & $-$24.47 \\
F2MS0921$+$1918 & 1.791 & 0.81 $\pm$ 0.15 & ...            & $-$29.13 \\
F2MS0932$+$3854 & 0.506 & 0.85 $\pm$ 0.12 & 1000 $\pm$ 150 & $-$27.08 \\
F2MS0943$+$5417 & 2.224 & 0.97 $\pm$ 0.16 & ...            & $-$31.07 \\
F2MS1004$+$1229 & 2.658 & 0.76 $\pm$ 0.24 & ...            & $-$31.61 \\
F2MS1005$+$6357 & 1.280 & 1.07 $\pm$ 0.11 & ...            & $-$28.31 \\
F2MS1012$+$2825 & 0.937 & 0.82 $\pm$ 0.10 & 1500 $\pm$ 200 & $-$26.04 \\
F2MS1030$+$5806 & 0.504 & 0.99 $\pm$ 0.06 & 4800 $\pm$ 700 & $-$26.55 \\
F2MS1036$+$2828 & 1.760 & 0.61 $\pm$ 0.07 & ...            & $-$28.96 \\
F2MS1113$+$1244 & 0.681 & 1.41 $\pm$ 0.11 & 3600 $\pm$ 600 & $-$28.37 \\
F2MS1118$-$0033 & 0.686 & 0.85 $\pm$ 0.11 & 2100 $\pm$ 500 & $-$25.89 \\
F2MS1121$+$2132 & 0.834 & 0.33 $\pm$ 0.11 & 1550 $\pm$ 400 & $-$23.88 \\
F2MS1151$+$5359 & 0.780 & 0.42 $\pm$ 0.08 & 3300 $\pm$ 500 & $-$24.13 \\
F2MS1243$+$1023 & 0.680 & 0.83 $\pm$ 0.08 & 1150 $\pm$ 200 & $-$25.77 \\
F2MS1248$+$0531 & 0.740 & 0.26 $\pm$ 0.07 & 1250 $\pm$ 200 & $-$23.45 \\
F2MS1256$-$0200 & 0.835 & 1.31 $\pm$ 0.09 & 3800 $\pm$ 500 & $-$27.92 \\
F2MS1307$+$2338 & 0.275 & 0.75 $\pm$ 0.09 & 1200 $\pm$ 200 & $-$23.17 \\
F2MS1309$+$6042 & 0.641 & 0.95 $\pm$ 0.13 & 2000 $\pm$ 400 & $-$24.87 \\
F2MS1313$+$1453 & 0.584 & 1.12 $\pm$ 0.15 & 3100 $\pm$ 500 & $-$26.59 \\
F2MS1341$+$3301 & 1.720 & 0.68 $\pm$ 0.09 & ...            & $-$28.78 \\
F2MS1344$+$2839 & 1.770 & 0.55 $\pm$ 0.12 & ...            & $-$28.82 \\
F2MS1353$+$3657 & 1.311 & 1.51 $\pm$ 0.16 & ...            & $-$29.81 \\
F2MS1415$+$0903 & 0.654 & 0.45 $\pm$ 0.11 & 4300 $\pm$ 500 & $-$24.10 \\
F2MS1434$+$0935 & 0.770 & 1.14 $\pm$ 0.10 & 2200 $\pm$ 400 & $-$27.20 \\
F2MS1456$+$0114 & 2.378 & 0.45 $\pm$ 0.16 & ...            & $-$29.24 \\
F2MS1500$+$0231 & 1.501 & 0.72 $\pm$ 0.11 & ...            & $-$28.44 \\
F2MS1507$+$3129 & 0.988 & 0.74 $\pm$ 0.09 & 1700 $\pm$ 250 & $-$26.55 \\
F2MS1523$+$0030 & 0.560 & 0.35 $\pm$ 0.10 & 1950 $\pm$ 200 & $-$23.24 \\
F2MS1540$+$4923 & 0.696 & 0.97 $\pm$ 0.10 & 3500 $\pm$ 300 & $-$26.46 \\
F2MS1548$+$0913 & 1.524 & 0.72 $\pm$ 0.06 & ...            & $-$28.36 \\
F2MS1549$+$0306 & 0.679 & 0.64 $\pm$ 0.08 & 3600 $\pm$ 500 & $-$24.97 \\
F2MS1551$+$0844 & 2.505 & 0.48 $\pm$ 0.18 & ...            & $-$29.76 \\
F2MS1615$+$0318 & 0.424 & 0.42 $\pm$ 0.06 & 3400 $\pm$ 300 & $-$23.25 \\
F2MS1618$+$3502 & 0.446 & 0.62 $\pm$ 0.20 & 700*           & $-$24.87 \\
F2MS1638$+$2707 & 1.692 & 0.64 $\pm$ 0.14 & ...            & $-$27.76 \\
F2MS1650$+$3242 & 1.059 & 0.79 $\pm$ 0.06 & 3800 $\pm$ 500 & $-$26.93 \\
F2MS1656$+$3821 & 0.732 & 0.88 $\pm$ 0.16 & 350*           & $-$25.14 \\
F2MS1708$+$2227 & 0.377 & 1.26 $\pm$ 0.12 & 3100 $\pm$ 700 & $-$24.69 \\
F2MS1720$+$6156 & 0.727 & 1.50 $\pm$ 0.09 & 3900 $\pm$ 400 & $-$27.54 \\
F2MS2152$-$0028 & 0.867 & 0.86 $\pm$ 0.10 & 2200 $\pm$ 200 & $-$25.77 \\
F2MS2339$-$0912 & 0.660 & 1.25 $\pm$ 0.06 & 2500 $\pm$ 300 & $-$28.29 \\
\enddata
\end{deluxetable*}

\begin{figure}[t]
\begin{center}
\plotone{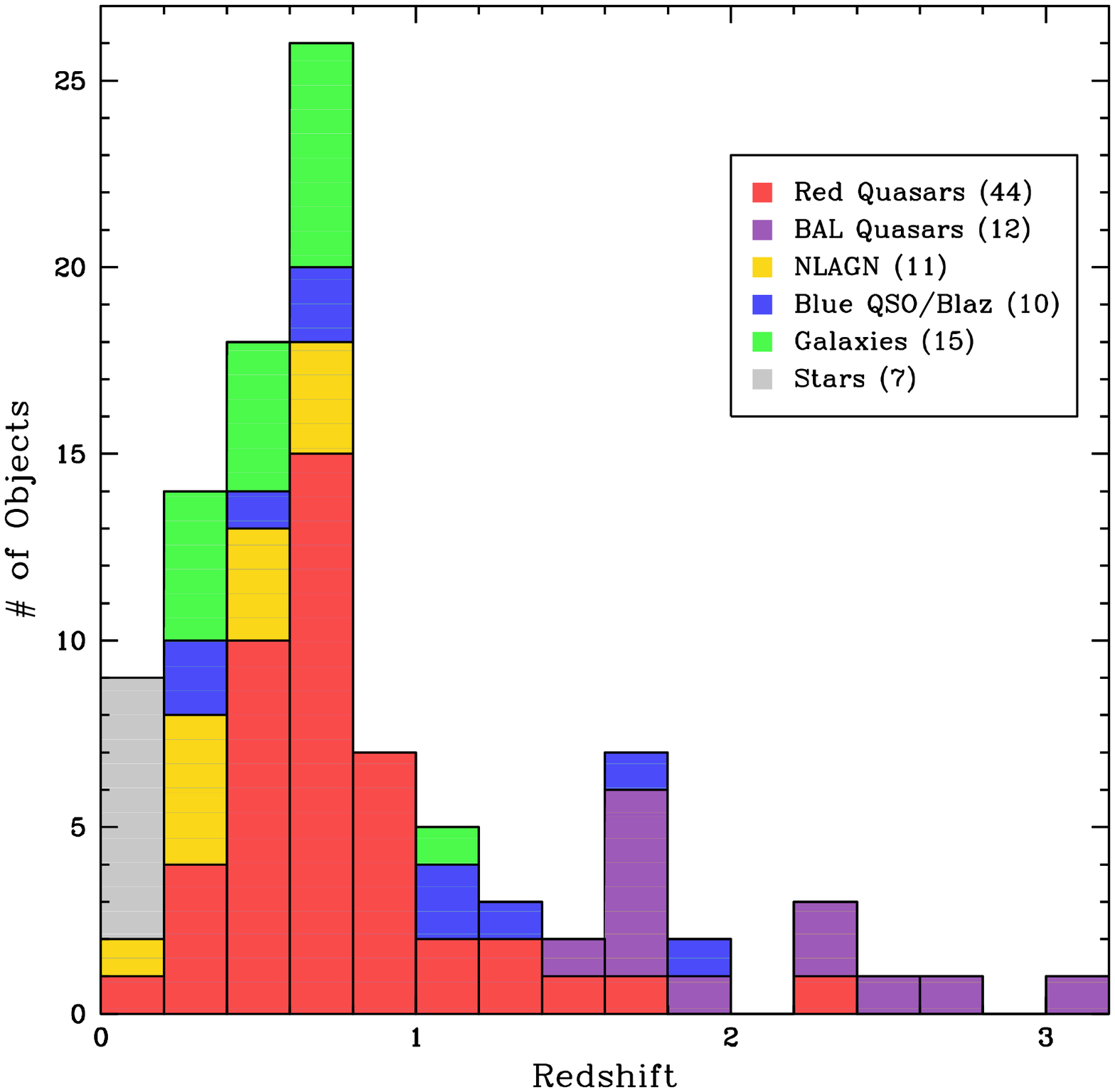}
\end{center}
\caption{Redshift distribution of the spectroscopically identified objects. 
Note the high fraction of BAL quasars at high redshift.
\label{redshift}}
\end{figure}

The optical identification of the F2MS objects yielded the following 
classifications: 56 red quasars, among them twelve Broad Absorption Line 
Quasars (BALs), eleven Narrow Line AGN (NLAGN), 15 Galaxies (either early type 
or starburst), ten BL Lacs or blue unabsorbed quasars and seven M-Stars. Their 
redshift distribution is shown in Figure \ref{redshift}. The redshift 
distribution, especially of the quasars, peaks around $z \sim$ 0.7. This is 
mostly a selection effect. At lower redshifts a large number of low luminosity 
object will not have been selected because the dominance of the host galaxy 
light prevents the quasar from making our color cuts. Host galaxies, even if 
red don't show extreme r'-K $>$ 5 colors so even if the quasar nucleus is very 
red the host galaxy will dilute the quasar's color contribution. At higher 
redshifts the relatively shallow 2MASS magnitude limit means that only the 
most luminous objects are selected.

\subsection{Properties of the Red Quasars}

Once we have derived the reddening of the quasars, we can also obtain their 
intrinsic luminosities ($M_i$) with the optical i magnitude corrected for 
obscuration. They are presented in Table \ref{spectra}. We use the 
K-correction for quasars in i-Band provided by \cite{Richards06}, but use 
their adjustment to $z=0$ ($M_i(z=0) = M_i(z=2)  + 0.596$). The F2MS red 
quasars all lie above the Quasar/Seyfert divide of $M_i = -22.5$ 
\citep{Richards06,hao05}, we are dealing with a very luminous sample. Three 
quasars have absolute magnitudes $<$ -30. These are known gravitational lenses 
whose computed luminosity is overestimated as a result of the lens 
magnification.

\begin{figure}[t]
\begin{center}
\plotone{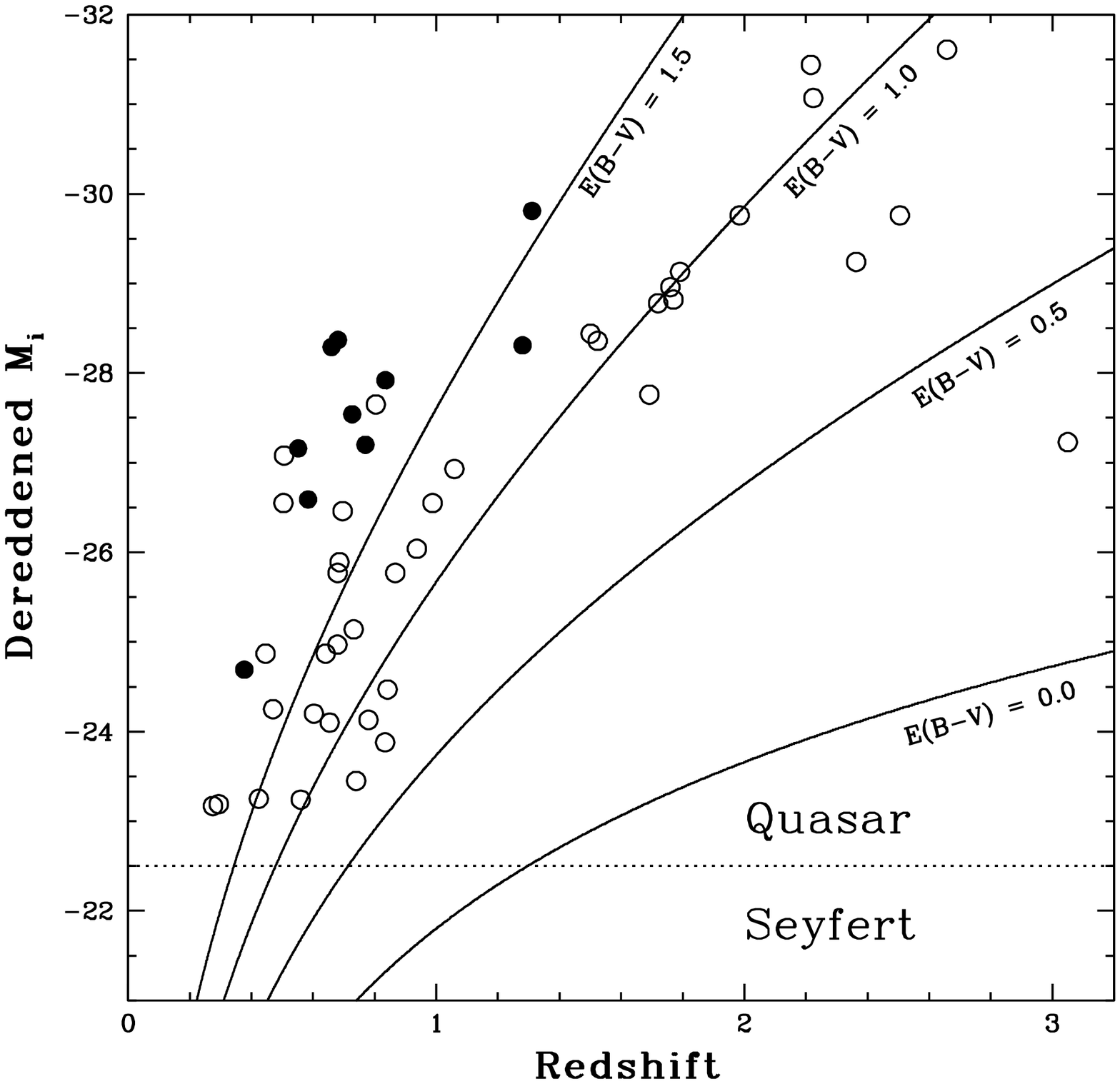}
\end{center}
\caption{Redness corrected absolute magnitude ($M_i$) of the F2MS red quasars 
versus redshift. The quasars with the highest reddenings ($E(B-V) > 1.0$) are 
indicated as filled dots. The curves show SDSS i $<$ 22.3 detection limit as a 
function of reddening $E(B-V)$. Note that at $z > 2.0$ quasars with such large 
reddenings become undetectable, which tells us that we must be missing a large 
fraction of objects at lower luminosities and higher redshifts (also see 
\cite{postman}).
\label{luminz}}
\end{figure}

Figure \ref{luminz} plots the absolute magnitude $M_i$ vs. redshift. The most 
obscured objects ($E(B-V) >$ 1.0) are shown as filled dots. Just as 
\cite{postman} concludes, we can only detect the most obscured objects 
when they are so intrinsically luminous and/or when they lie at a 
relatively low redshift. We are likely only detecting the most luminous tip of 
the red quasar iceberg. As expected, the highest redshift objects also have 
the highest absolute magnitudes. However, at low redshift there are not that 
many low luminosity QSOs or obscured Seyferts. This can be a selection effect 
because the AGN component for Seyfert galaxies is so diluted by the host 
galaxy light, so that its colors are not quasar-like (e.g. \cite{satyapal}). 
However, the steep rise in luminosity in Figure \ref{luminz} also brings up 
the question of whether they have the same luminosity function as unobscured 
quasars.

At first it would seem contradictory to claim that they do not, yet if we 
believe that dust-reddened quasars are in a young dust-enshrouded 
evolutionary phase, this phase would be highly intrinsically luminous as the 
central engine is accreting at high efficiencies \citep{hop06}. Some 
{\it Spitzer} selected red QSOs also are only found at very high intrinsic 
luminosities \citep{polletta2}. Yet the results are not very conclusive and 
there are counterexamples to this claim \citep{lacy07}.

\begin{figure}
\begin{center}
\plotone{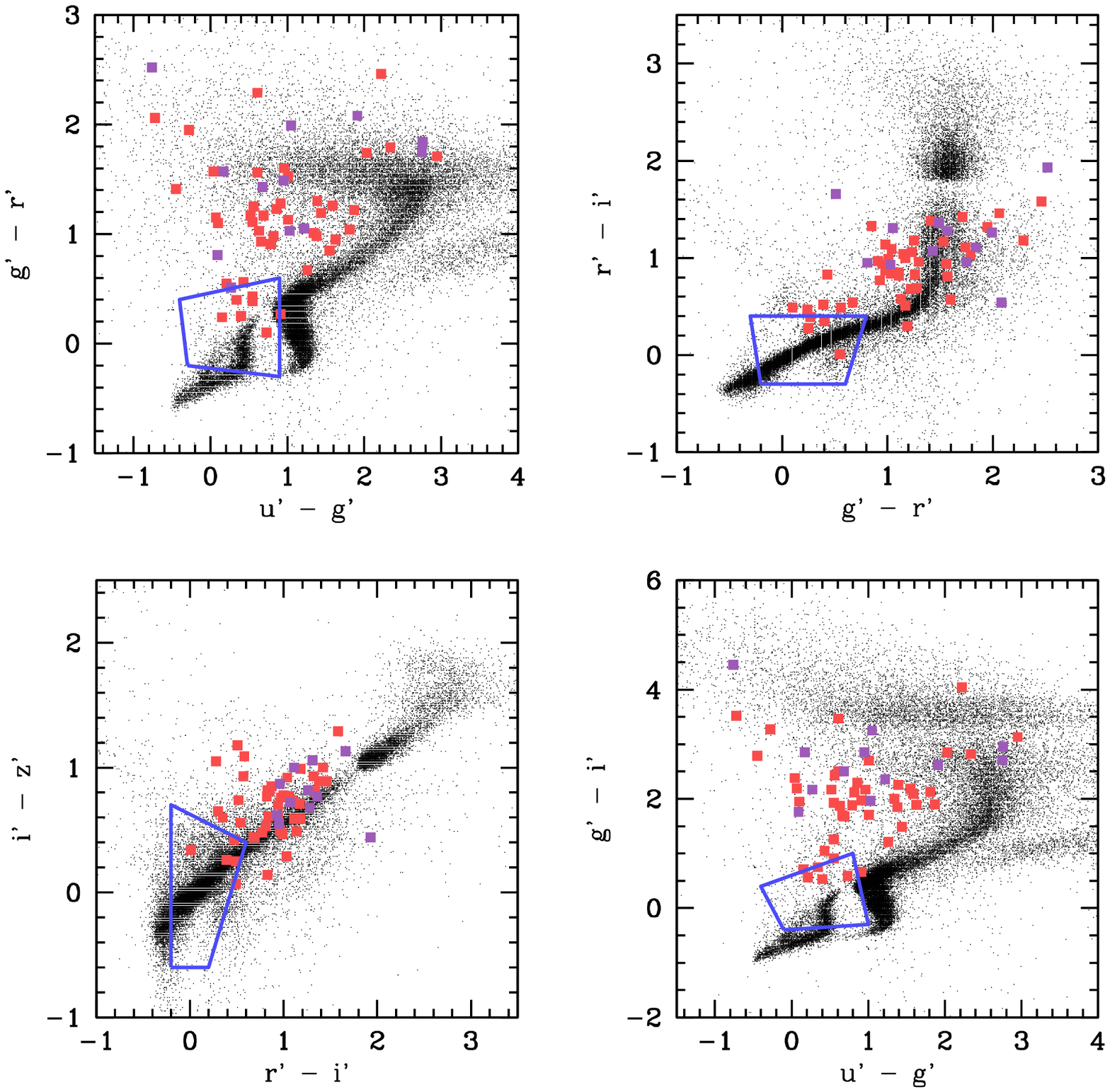}
\end{center}
\caption{Various optical colors of the objects classified as Red Quasars and 
BAL Quasars. The small black dots are the colors of stars in SDSS DR2. The 
blue boxes are the representation of the quasars contours of the SDSS z $<$ 3 
from \cite{richards04}. Note that most of our Red Quasars and {\it all} of 
the BAL quasars lie outside those boxes and that the Red Quasars often fall 
into the stellar locus making it difficult for typical quasar search 
algorithms to find these objects.
\label{colors}}
\end{figure}

Most of our sources have magnitudes that fall below the normal cutoff 
(i=19.1) for spectroscopy with SDSS. Some of them were targeted for 
spectroscopy as potential high redshift quasars or serendipitous radio source 
identifications. Of our 122 targets 49 (40\%) had flags that would select 
them for potential QSO target follow up. Therefore, optical surveys such as 
the SDSS would miss over 60\% of these objects with traditional color 
selection algorithms. \cite{richards04} have undertaken a project to determine 
if even fainter parts of the color space are suitable for finding quasars and 
they have potentially found over 100,000 QSOs in this manner. Figure 
\ref{colors} presents the SDSS colors of our red QSOs, plotted with red 
squares, the BALQSOs are denoted with a purple square. The small black dots 
denote the colors of the DR2 stars taken from the SDSS website. The blue box 
in the image is representative of the peak contour from \cite{richards04}. 
Note that most of the F2MS Red Quasars and {\it all} of the F2MS BALQSOs lie 
outside that box. Furthermore, many of the F2MS red quasars lie on the stellar 
locus for different color-color plots, making optical selection algorithms 
unsuitable for identifying these objects. To include these objects, one needs 
near- and mid-Infrared photometry, which is insensitive to this amount of dust 
reddening (e.g. \cite{lacy,stern}). These optical selection caveats are taken 
into account for deep fields such as the Spitzer FLS survey, which has 
multiwavelength observations \citep{Richardsaas}.

\subsection{An anomalously high fraction of Broad Absorption Line Quasars}

A BAL quasar is defined to be any quasar that shows significant broad 
absorption blueward of various quasar emission lines, such as MgII 
$\lambda$2800 or CIV $\lambda$1549. BALs usually can only be identified at the 
high redshifts where the absorption troughs, usually present in the restframe 
UV part of the spectrum, are redshifted into the optical regime. For LoBALs, 
with their absorption troughs bluewards of the MgII line, this requires a 
redshift range of 0.9 $<$ z $<$ 2.6 for the typical ESI spectral coverage of 
4000 - 10,000 \AA. While there are some lower redshift objects which show the 
MgII emission line, ESI is not optimal in those wavelength regimes to 
determine absorption troughs; there is an Echelle break around 4500 \AA and 
below 5000 \AA the error spectrum rises steeply. LoBALs can also be identified 
by their AlIII $\lambda$1859 absorption at higher redshifts. For HiBALs, which 
only show prominent absorption troughs in the high ionization lines such as 
CIV, we usually need redshifts of z $>$ 1.7; although even at these redshifts 
the flux in the blue part of the spectrum is suppressed as our objects have 
large dust-reddenings. For the F2MS sample, 21 red quasars fall into the 
redshift range where BALs can be identified. 

The definition of the Balnicity index was selected so that host and 
intervening narrow absorption systems are not selected erroneously. The 
traditional Balnicity index is defined by \cite{weymann} as:

\begin{equation}
BI = \int_{3,000}^{25,000} \left[ 1 - \frac{f(\mu)}{0.9} \right] C d\mu
\end{equation}

\noindent
where $f(\mu)$ is the normalized flux, $C = 1$ at absorption through 
velocities more than 2000 km/s from the start of a continuous trough and zero 
elsewhere. Usually uncertainties in the continuum dominate over the formal 
error, so we do not include errors for our estimated Balnicities. 

\begin{figure*}
\begin{center}
\includegraphics[width=15cm]{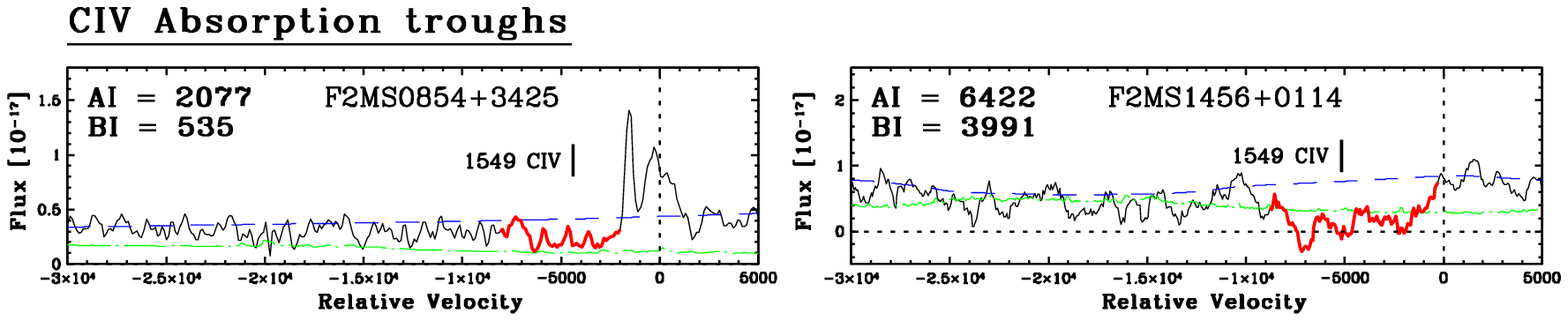}
\includegraphics[width=15cm]{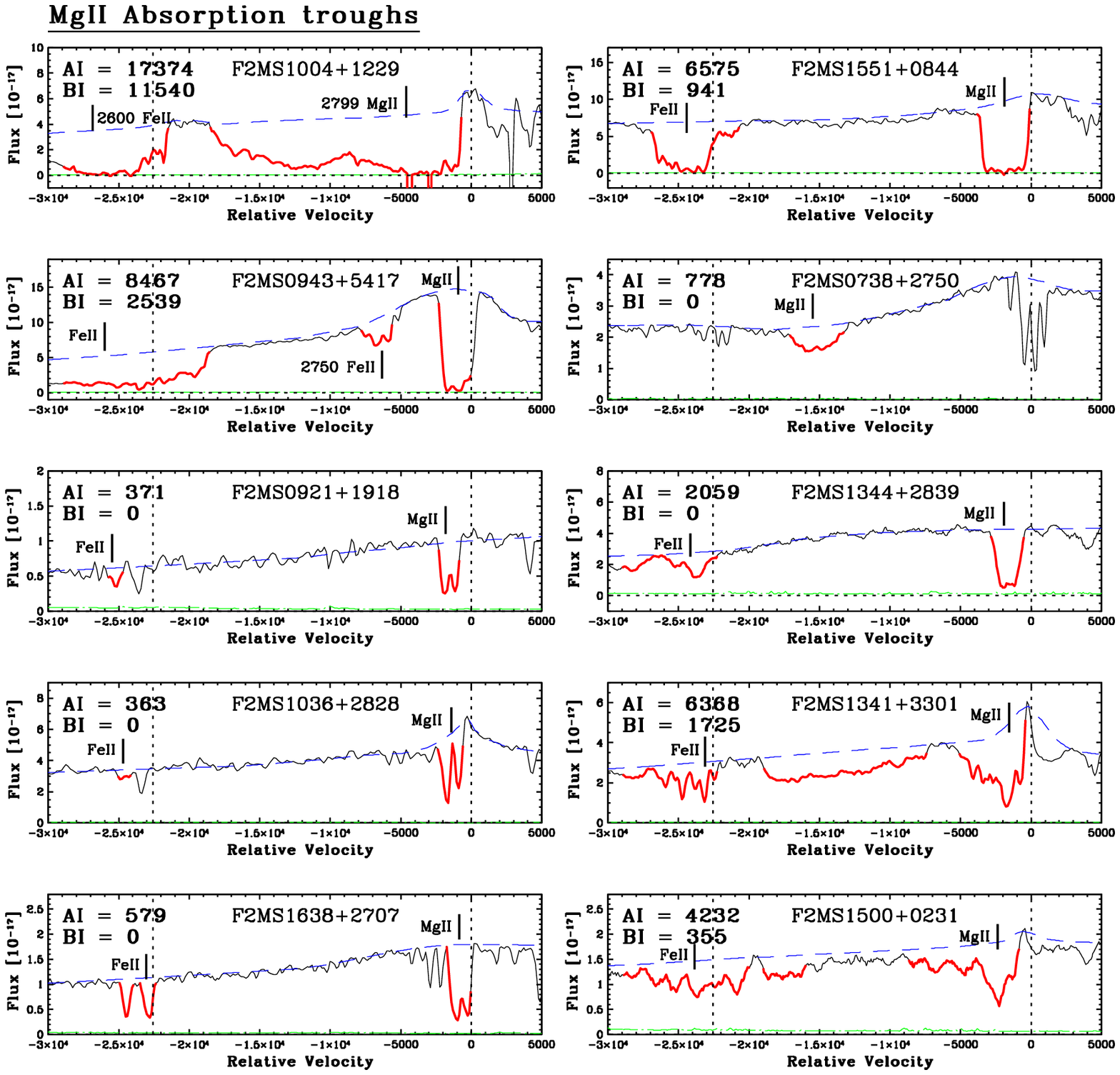}
\end{center}
\caption{\footnotesize Absorption troughs for the BAL and mini-BAL objects in 
the F2MS sample at $z > 0.9$. We plot the velocity range between -30,000 km/s 
and 5000 km/s of the emission redshift, which is the typical range for 
identification using the Balnicity and Absorption Indices. The two upper 
spectra show the CIV absorption troughs, where MgII had no spectral coverage, 
while the lower panels show the troughs for the MgII absorption troughs, which 
had higher signal to noise. The black line is the non-normalized spectrum, 
with the bold red parts marked as the absorption troughs identified through 
the Absorption Index. The blue dashed line indicates the continuum used to 
obtain the normalized spectrum and the green line represents the error 
spectrum. In the panels showing MgII absorption troughs, we also additionally 
denote the prominent FeII absorption feature at 2600 \AA. The objects with 
Balnicity Indices above zero all show absorption at in the FeII complex, 
hinting that they belong to the rare class of FeLoBALs. \label{balplot}}
\end{figure*}

With somewhat better resolution data and better continuum fitting techniques, 
\cite{Trump} decided to use the Absorption Index to identify BALQSOs. The 
Absorption Index is defined by \cite{ha02} as

\begin{equation}
AI = \int_0^{29,000} \left[ 1 - f(\mu) \right] C' d\mu
\end{equation}

\noindent
where $f(\mu)$ is the normalized flux. $C' = 0$ except in continuous troughs 
with the minimum width of 1000 km/s and a minimum 10\% depth, in which case 
$C' =- 1$.  

There is still disagreement over the exact definition of a BAL. While the 
traditional, more conservative Balnicity index measure ensures that the 
absorption is from a nuclear outflow, and effectively excludes ``associated 
absorbers'', this definition could potentially exclude unusual or interesting 
BALs (e.g. \cite{bobbal}). The Absorption Index, on the other hand, allows for 
somewhat narrower troughs and absorption at the emission redshift. 
\cite{knigge} warn that a BAL classification via the Absorption index needs to 
be reconsidered, since its distribution is bimodal \cite{knigge}. \cite{Trump} 
find that 26\% of all QSOs are BALs if defined by the absorption index. This 
is higher than the $\sim$12\% fraction determined using the traditional 
Balnicity Index.

We fitted the continuum for the 21 red F2MS quasars that were in the redshift 
range suited to be identified as BALs. Our small sample size allowed us to 
adjust the continuum at the emission lines by hand rather than let an 
automated process run over it. We then normalized the spectra and calculated 
the Balnicity and Absorption Indices according to the formulas above. We used 
mostly the MgII line as a proxy for balnicity, since the CIV line was very 
noisy for the objects with $z < 2$.

\begin{figure*}
\begin{center}
\plotone{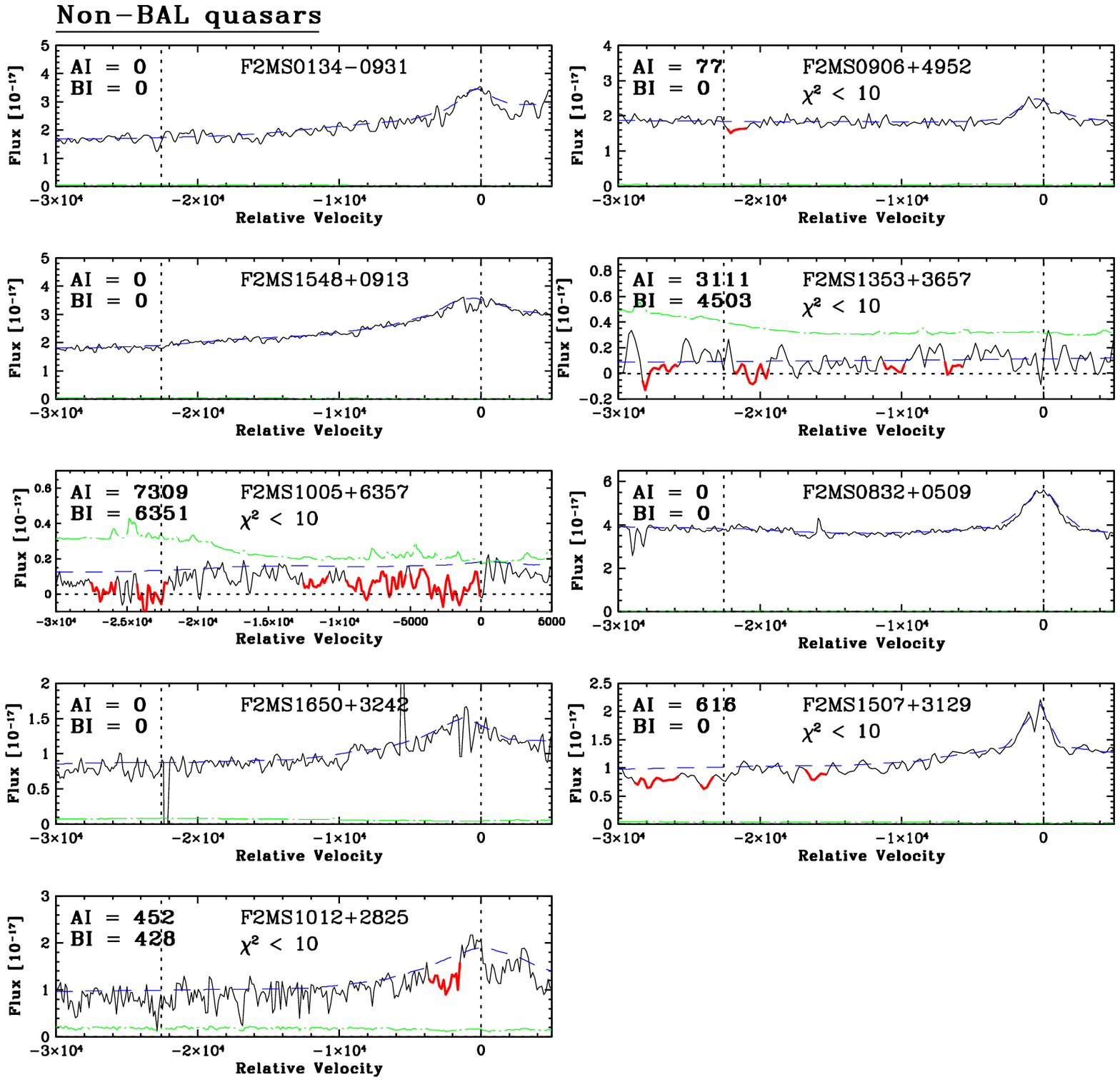}
\caption{Same range and properties as Figure \ref{balplot}, only for the 
non-BAL quasars at $z > 0.9$. The two dotted lines represent where blueward 
the MgII and FeII absorption features would lie. F2MS1353$+$3657 and 
F2MS1005$+$0231 are clearly dominated by noise, so we don't include them in 
our calculation for the BAL fraction.\label{nonbals}}
\end{center}
\end{figure*}

As mentioned before, the ESI spectrograph has a poor throughput at the blue 
end of the spectrum. Not only are the quasars very extinguished in that part, 
but the poor throughput in the near-UV from ESI makes the spectrum very noisy. 
It is very difficult to distinguish absorption MgII troughs qualitatively 
below a redshift of 0.9. 

We also introduce the reduced chi-squared of the trough as defined by 
\cite{Trump}:

\begin{equation}
\chi^2_{trough} = \sum \frac{1}{N} \left[ \frac{1 - f(\nu)}{\sigma} \right]^2
\end{equation}

N is the number of pixels in the trough and $f(\nu)$ is the normalized 
template-subtracted flux. The higher $\chi^2_{trough}$, the less likely the 
trough is due to noise or a really shallow artificial trough introduced by a 
bad continuum fit. We require a $\chi^2_{through} > 10$ to be considered true 
BAL troughs. We make one exception for F2MS1456$+$0114, whose CIV line trough 
has a $\chi^2_{trough} < 10$, but shows such prominent FeII absorption, that 
it is also very likely to have very high MgII absorption if it had spectral 
coverage there. For the calculation of the MgII fraction, we do not include 
F2MS1353$+$3657 and F2MS1005$+$6357, which clearly have noise dominated 
spectra, where the error is sometimes double that of the signal. We do however 
include F2MS0906$+$4952, F2MS1507$+$3129 and F2MS1012$+$2825 as non-BALs; from 
visual inspection, we can deduce that the troughs are not real. However, we do 
note that there is generally a lack of BALs in our sample below redshift $z$ = 
1.5. We don't know if that is a selection effect or if there is are two 
different populations at high and low redshift.

12 of the 19 (63\%) eligible objects were identified as BALQSOs using the AI. 
However, five of the objects would have been rejected as a BAL by 
\cite{weymann} since the absorption troughs were either too narrow (below 
2000 km/s, but above 1000 km/s) or because they were so close to the emission 
redshift and could have been mistaken with absorbers associated with the host 
galaxy (although most host galaxies do not have such large velocity 
dispersions). We label these five objects as Mini-BALs, yet include them in 
the discussion about BALs. If we were to exclude the objects with zero BIs, 
seven out of 19 F2MS quasars within the appropriate redshift are BALs (37\%). 
Figure \ref{balplot} shows the spectra of the BALs with the absorption troughs 
bolded in red. Non-BALs for a redshift $>$ 0.9 are shown in Figure 
\ref{nonbals}, with the noisy spectra included for completeness.

\begin{deluxetable*}{lccccl}
\tabletypesize\footnotesize
\tablecaption{Absorption trough properties of Quasars with $z > 0.9$ \label{bal}}
\tablehead{
\colhead{Source} & \colhead{Redshift} & \colhead{Balnicity} & 
\colhead{Absorption Index} & \colhead{$\chi^2_{through}$} & \colhead{Type}}
\startdata
F2MS0854$+$3425 & 3.050 & 535$^a$  & 2077$^a$ & 11.65 & BAL / HiBAL (no MgII) \\
F2MS1004$+$1229 & 2.668 & 11540    & 17374    &  3062 & BAL / FeLoBAL \\
F2MS1551$+$0844 & 2.505 & 941      & 6575     & 28.90 & BAL / FeLoBAL \\
F2MS1456$+$0114 & 2.378 & 3991$^a$ & 6422$^a$ & 4.491 & BAL / FeLoBAL (based on FeII) \\
F2MS0943$+$5417 & 2.224 & 2539     & 8467     &  2231 & BAL / FeLoBAL \\
F2MS0134$-$0931 & 2.216 & 0        & 0        & -     & non-BAL \\
F2MS0738$+$2750 & 1.985 & 0        & 778      & 83.78 & Mini-Bal / LoBAL \\
F2MS0921$+$1918 & 1.791 & 0        & 317      & 49.59 & Mini-Bal / FeLoBAL \\
F2MS1344$+$2839 & 1.770 & 0        & 2059     & 12.39 & Mini-BAL / FeLoBAL \\
F2MS1036$+$2828 & 1.761 & 0        & 363      &  1241 & Mini-BAL / FeLoBAL \\
F2MS1341$+$3301 & 1.720 & 1725     & 6368     &  1031 & BAL / FeLoBAL \\
F2MS1638$+$2707 & 1.692 & 0        & 579      & 217.1 & Mini-BAL / FeLoBAL \\
F2MS0906$+$4952 & 1.635 & 0        & 77       & 1.421 & non-BAL \\
F2MS1548$+$0913 & 1.524 & 0        & 0        & -     & non-BAL \\
F2MS1500$+$0231 & 1.511 & 355      & 4232     & 10.61 & BAL / FeLoBAL \\
F2MS1353$+$3657 & 1.312 & 4503     & 3111     & 1.166 & too noisy \\
F2MS1005$+$6357 & 1.282 & 6351     & 7309     & 4.512 & too noisy \\
F2MS0832$+$0509 & 1.070 & 0        & 0        & -     & non-BAL \\
F2MS1650$+$3241 & 1.059 & 0        & 0        & -     & non-BAL \\
F2MS1507$+$3129 & 0.988 & 0        & 616      & 6.933 & non-BAL \\
F2MS1012$+$2825 & 0.937 & 428      & 452      & 0.611 & non-BAL \\
\enddata
\tablecomments{$^a$ CIV derived Balnicities and Absorption Indices, all other 
are for the MgII line, which in most cases had much higher signal to noise in 
the spectra}
\end{deluxetable*}

Our conservative BAL fraction of 37\% alone is already much higher than 
almost all other quasar surveys. But even more interesting is the fact that 
{\it all} seven BALQSOs that had MgII wavelength coverage were LoBALs. 
F2MS0854$+$3425 has no MgII coverage, but shows no AlIII $\lambda$1859 
absorption through, so it is most likely a HiBAL. Also remarkable is that 
{\it all} of the objects classified as LoBALs also were of the rare population 
of FeLoBALs. Our conservative LoBAL and also FeLoBAL fraction is 32\% (6/19). 
\cite{Trump} finds LoBALs and FeLoBALs fraction to be 0.5\% and 0.3\% 
respectively. This two order of magnitude discrepancy for (Fe)LoBAL fraction 
is truly remarkable; dust-absorbed and reddened quasars show a much higher 
than usual fraction of LoBALs than quasars selected in optical surveys.

\section{Discussion}

We have expanded the original FIRST-2MASS red quasar survey to include the 
whole extent of the 2MASS point source catalog but only include objects which 
have a very red color (r'-K $>$ 5). This yielded 122 red quasar candidates. 
Spectroscopic follow-up on 100 of the sources revealed that well over 50\% are 
red Type-1 QSOs, consistent with the original FIRST-2MASS Red Quasar survey.

As described in Section \ref{specobs} the deduced reddening is most likely 
due to dust absorption. The question remains where the obscuring dust is 
located. It could be in regions of the torus, where sublimation does not 
affect it, or it could be in a cold outside absorber, i.e. the host galaxy. 
Some clues as to the nature of the extinction in these objects came from our 
HST/ACS study of a sample of 13 F2M quasars (those in the F2MS sample are 
labeled in Table \ref{observe}). The images showed impressive evidence 
for a recent merger in an unusually high percentage (85\%) of the red quasars. 
Also, there was a weak correlation between the reddening and the amount of 
interaction displayed in the host galaxy measured by the Gini coefficient. 
These dust-reddened quasars may be young, recently ignited objects, where 
quasar feedback has yet to quench the star formation activity 
\citep{redqso-hst}. Ongoing Spitzer IRS and MIPS observations of the HST/ACS 
sample should help to test this hypothesis, as we will be able to measure PAH 
strengths and 70$\mu$m dust emission likely associated with young star forming 
host galaxies. The Spitzer IRS spectra will also describe the absorbing dust 
properties better by analyzing the 9.7 $\mu$m Silicate absorption feature and 
relating that to our deduced reddening. 

There is still debate over how biased our surveys are. Radio-selected samples 
tend to have redder colors. However, IR selected samples tend to have larger 
fractions of BALQSOs \citep{dai}. There has been some debate over whether 
selection biases in the optical make one underestimate the fraction of BALQSOs 
\citep{krolik}. Essentially none of the BALs discovered with the F2MS survey 
would have been discovered with the SDSS color selection techniques (except 
for serendipitous spectra or searches for high redshift quasars). Hence, our 
deduced fraction of BALs is too high (63\%) and even the conservative fraction 
of (40\%) deduced by \cite{dai} is quite large. It becoms difficult to 
explain these nature objects as a pure orientation effect as the wind opening 
angle would be very large. However, since we are biased in only looking at 
dust-reddened objects, the dust could be present in the BAL clouds themselves 
and we would find naturally only find such a large number of BALs in objects 
that are reddened. Also, in the orientation picture, the highly collimated BAL 
winds could be grazing the dust torus, meaning that they are naturally redder 
than usual. These caveats aside, we believe that with such large dust 
reddenings, the evidence for mergers at lower redshift for similarly selected 
objects, and the large intrinsic fraction lead us to believe that BALQSOs are 
in an early evolutionary phase in the lifetime of a quasar. Their powerful 
winds might be the quenchers of star formation in the host galaxy, often 
invoked as quasar feedback (e.g. \cite{sr98,kauffmann,li07}).

The ``youth hypothesis'' for LoBALs fits very well with the picture of 
dust-reddened quasars being an evolutionary step in the lifetime of a quasar. 
The results from the hydrodynamic simulations predict that there is some 
amount of quasar feedback seen as a kind of ``blowout'' just after the phase 
where the merger has coalesced and the quasar is starting to ignite (albeit, 
enshrouded in dust) \citep{hop07}. Whether this quasar feedback is actually 
done by a quasar wind that produces BAL-Type spectra is not clear yet though 
there is some evidence that LoBAL winds can extend into the host galaxy 
\citep{dekool}. Recent poststarburst galaxies also show evidence for massive 
outflows which must have originated in a quasar when the starburst was still 
young \citep{tremonti}. 

\acknowledgments

The authors wish to thank Bryn Feldman for help with the first selections of 
this catalog and for help in carrying out the observations at Lick 3m 
telescope. We are also thankful for the helpful comments from an anonymous 
referee on the paper. This work was partly performed under the auspices of the 
US Department of Energy by Lawrence Livermore National Laboratory under 
contract No. DE-AC52-07NA27344.

This publication makes use of data products from the Two Micron All Sky 
Survey, which is a joint project of the University of Massachusetts and the 
Infrared Processing and Analysis Center/California Institute of Technology, 
funded by the National Aeronautics and Space Administration and the National 
Science Foundation

Funding for the Sloan Digital Sky Survey (SDSS) has been provided by the 
Alfred P. Sloan Foundation, the Participating Institutions, the National 
Aeronautics and Space Administration, the National Science Foundation, the 
US Department of Energy, the Japanese Monbukagakusho and the Max Planck 
Society.


\clearpage

\begin{landscape}

\begin{deluxetable}{lllrrccccccccccllcll}
\tabletypesize\tiny
\tablewidth{0pt}
\tablecaption{Properties and identification of red quasar candidates \label{observe}}
\tablehead{
\colhead{No} & \colhead{R.A.} & \colhead{Dec} & \colhead{$F_{pk}$} & 
\colhead{u'} & \colhead{g'} &\colhead{r'} & \colhead{i'} &  \colhead{J} & 
\colhead{K} & \colhead{r'$-$K} & \colhead{J$-$K} & \colhead{Spect.} & 
\colhead{Type} & \colhead{$z$} & \colhead{comments} & \colhead{Reference}}
\startdata  
001 & 00:36:59.85 & $-$01:13:32.30 &   1.92 & 22.09 & 21.40 & 20.23 & 19.72 & 16.57 & 13.63 & 6.60 & 2.94 & ESI    & QSO     & 0.294  & galaxy group in foreground,\citetalias{f2m07} & \\
002 & 01:34:35.68 & $-$09:31:03.00 & 900.38 & 25.87 & 23.65 & 21.19 & 19.61 & 16.19 & 13.58 & 7.61 & 2.61 & NOT    & QSO     & 2.216  & gravitational lens,\citetalias{f2m07} & Gr02,Wi02,+33 \\
003 & 01:56:47.61 & $-$00:58:07.46 &   6.51 & 21.38 & 20.92 & 20.11 & 19.54 & 17.43 & 14.87 & 5.24 & 2.56 & ESI    & NLAGN   & 0.507  & & \\
004 & 07:30:51.03 & $+$25:38:59.03 &   1.23 & 22.01 & 20.59 & 19.49 & 19.07 & 16.85 & 14.15 & 5.34 & 2.70 & ESI    & NLAGN   & 0.289  & \citetalias{f2m07} & \\  
005 & 07:36:24.18 & $+$42:22:17.15 &   3.93 & 22.50 & 22.34 & 21.63 & 20.90 & 17.08 & 15.46 & 6.17 & 1.62 & none   & -       & -      & & \\
006 & 07:38:05.95 & $+$29:51:09.90 &   2.64 & 20.89 & 22.12 & 20.82 & 19.15 & 16.94 & 15.28 & 5.54 & 1.66 & SDSS   & Galaxy  & 0.393  & \citetalias{f2m07} & \\
007 & 07:38:20.10 & $+$27:50:45.49 &   2.64 & 23.78 & 22.73 & 20.74 & 19.48 & 17.05 & 15.28 & 5.46 & 1.77 & ESI    & QSO     & 1.985  & Mini-BAL,\citetalias{f2m07} & Gr02,Ri03,+4 \\
008 & 07:55:01.09 & $+$40:16:32.16 &   1.22 & 22.68 & 21.71 & 20.22 & 19.58 & 17.43 & 14.95 & 5.27 & 2.48 & SDSS   & Galaxy  & 0.381  & & \\
009 & 08:01:08.26 & $+$44:01:09.98 & 211.67 & 21.03 & 20.37 & 19.84 & 19.44 & 16.45 & 14.83 & 5.01 & 1.62 & SDSS   & BL Lac? & 1.072? & featureless spectrum & \\
010 & 08:04:56.87 & $+$19:17:26.85 &   5.32 & 24.25 & 23.41 & 21.84 & 20.48 & 17.50 & 15.53 & 6.31 & 1.97 & none   & -       & -      & & \\ 
011 & 08:07:37.42 & $+$44:36:43.42 &   1.27 & 23.87 & 22.40 & 20.80 & 19.55 & 17.07 & 15.49 & 5.31 & 1.58 & ESI    & Galaxy  & 0.665  & low S/N & \\
012 & 08:13:07.44 & $+$33:00:04.11 &   1.62 & 23.74 & 25.11 & 20.72 & 19.02 & 16.81 & 15.24 & 5.48 & 1.57 & SDSS   & Galaxy  & 0.393  & & \\
013 & 08:15:25.26 & $+$21:01:40.91 &   2.42 & 25.86 & 22.42 & 20.53 & 19.46 & 17.40 & 15.45 & 5.08 & 1.95 & none   & -       & -      & high proper motion star? & Kh04,+1\\
014 & 08:25:02.05 & $+$47:16:51.96 &  61.12 & 21.57 & 21.02 & 20.59 & 19.76 & 17.15 & 14.17 & 6.42 & 2.98 & ESI    & QSO     & 0.804  & HST image,\citetalias{f2m07} & Le01,VV06,+3 \\ 
015 & 08:32:11.64 & $+$05:09:01.04 &  42.17 & 21.32 & 21.10 & 20.55 & 20.54 & 16.51 & 14.73 & 5.82 & 1.78 & ESI    & QSO     & 1.070  & not very red & Am88 \\
016 & 08:34:07.01 & $+$35:06:01.83 &   1.22 & 22.80 & 20.93 & 19.71 & 19.03 & 16.54 & 14.64 & 5.07 & 1.90 & ESI    & QSO     & 0.470  & HST image \citetalias{f2m07} & Sc05,VV06 \\
017 & 08:39:47.40 & $+$02:04:34.07 &   1.82 & 24.05 & 21.61 & 20.61 & 20.08 & 16.90 & 14.94 & 5.67 & 1.96 & ESI    & BL Lac? & ?      & featureless spectrum & \\
018 & 08:41:04.98 & $+$36:04:50.09 &   6.49 & 23.21 & 21.83 & 20.85 & 19.98 & 17.63 & 14.91 & 5.94 & 2.72 & ESI    & QSO     & 0.552  & HST image,\citetalias{f2m07} & Ur05,VV06 \\  
019 & 08:48:46.73 & $+$24:40:56.71 &   3.14 & 23.86 & 22.25 & 20.61 & 19.15 & 16.88 & 15.42 & 5.19 & 1.46 & none   & -       & -      & & \\
020 & 08:54:45.47 & $+$34:25:49.18 &   1.66 & 23.70 & 22.67 & 21.64 & 20.71 & 17.70 & 15.44 & 6.20 & 2.26 & ESI    & QSO     & 3.050  & BAL & \\
021 & 09:06:24.32 & $+$00:15:37.80 &   1.04 & 24.60 & 22.73 & 20.95 & 19.29 & 16.92 & 15.56 & 5.39 & 1.36 & ESI    & Star    & 0.000  & & \\
022 & 09:06:51.52 & $+$49:52:35.97 &  67.87 & 22.39 & 21.48 & 21.21 & 20.82 & 17.02 & 15.14 & 6.07 & 1.88 & ESI    & QSO     & 1.635  & not very red,\citetalias{f2m07} & Ur05,VV06 \\
023 & 09:11:57.55 & $+$01:43:27.77 &   4.49 & 22.95 & 22.13 & 21.15 & 20.01 & 16.46 & 14.87 & 6.28 & 1.59 & SDSS   & QSO     & 0.603  & very weak quasar & Za03,Sc03,+3 \\
024 & 09:15:01.71 & $+$24:18:12.24 &   9.80 & 20.94 & 20.54 & 20.29 & 20.01 & 16.56 & 13.82 & 6.47 & 2.74 & ESI    & QSO     & 0.842  & HST image, IRAS source,\citetalias{f2m07} & Ur05,Ba06 \\
025 & 09:21:45.69 & $+$19:18:12.64 &   4.40 & 22.73 & 22.46 & 21.95 & 20.29 & 16.76 & 14.58 & 7.37 & 2.18 & ESI    & QSO     & 1.791  & mini-BAL,\citetalias{f2m07} & Ki99,Sm02,VV06 \\
026 & 09:29:31.89 & $+$63:50:28.82 &   6.96 & 23.58 & 23.98 & 22.66 & 22.54 & 18.59 & 15.37 & 7.29 & 3.22 & ESI    & Galaxy  & 0.380  & Starburst & \\
027 & 09:32:33.29 & $+$38:54:28.12 &  40.92 & 22.32 & 21.53 & 20.62 & 19.65 & 17.75 & 15.14 & 5.48 & 2.61 & ESI    & QSO     & 0.506  & wrong reference? & Be96 \\
028 & 09:36:20.99 & $+$18:36:15.05 &   1.58 & 23.17 & 22.44 & 21.14 & 20.13 & 17.51 & 15.76 & 5.38 & 1.75 & none   & -       & -      & side arm of FRII? & \\
029 & 09:43:17.68 & $+$54:17:05.49 &   1.70 & 25.40 & 23.49 & 21.41 & 20.87 & 16.03 & 14.25 & 7.16 & 1.78 & ESI    & QSO     & 2.224  & impressive BAL & Sc05,VV06 \\
030 & 10:04:24.87 & $+$12:29:22.38 &  11.42 & 23.82 & 24.58 & 22.06 & 20.13 & 16.50 & 14.53 & 7.53 & 1.97 & ESI    & QSO     & 2.658  & gravitationally lensed BAL,\citetalias{f2m07} & Gr02,La02,+8 \\
031 & 10:05:42.83 & $+$63:57:23.69 &   1.82 & 24.62 & 24.01 & 21.72 & 20.54 & 17.29 & 15.40 & 6.32 & 1.89 & ESI    & QSO     & 1.280  & broad MgII? & \\
032 & 10:08:00.83 & $+$06:21:21.02 & 350.92 & 18.65 & 18.12 & 17.65 & 17.29 & 14.12 & 12.46 & 5.19 & 1.66 & ESI    & Blazar  & 1.720  & Highly variable & Sn02,Oj06,+6 \\
033 & 10:09:46.52 & $+$32:08:14.54 &   8.78 & 24.41 & 22.13 & 20.84 & 20.23 & 18.13 & 15.12 & 5.72 & 3.01 & none   & -       & -      & & \\
034 & 10:12:30.49 & $+$28:25:27.15 &   9.23 & 24.13 & 22.54 & 21.28 & 20.45 & 17.23 & 15.24 & 6.04 & 1.99 & ESI    & QSO     & 0.937  & HST image,\citetalias{f2m07} & Ur05,VV06 \\
035 & 10:30:39.62 & $+$58:06:11.41 & 113.33 & 21.70 & 20.31 & 19.01 & 18.05 & 15.99 & 13.83 & 5.18 & 2.16 & SDSS   & QSO     & 0.504  & & Sn01,Sc05,VV06 \\
036 & 10:36:33.54 & $+$28:28:21.58 &   4.25 & 21.18 & 21.09 & 20.28 & 19.33 & 16.99 & 15.25 & 5.03 & 1.74 & ESI    & QSO     & 1.760  & Mini-BAL,\citetalias{f2m07} & Ur05 \\
037 & 10:40:43.66 & $+$59:34:09.55 &   7.41 & 20.45 & 18.66 & 17.71 & 16.88 & 14.84 & 11.82 & 5.89 & 3.02 & ESI    & NLAGN   & 0.147  & IRAS source & deG92,Hu03,+5 \\
038 & 10:43:58.87 & $+$17:36:06.91 &   3.94 & 23.60 & 21.82 & 20.66 & 19.77 & 16.91 & 15.22 & 5.44 & 1.69 & none   & -       & -      & & \\
039 & 10:49:02.95 & $+$40:10:31.66 &   1.96 & 24.00 & 23.77 & 22.07 & 20.78 & 17.19 & 14.97 & 7.10 & 2.22 & ESI    & NLAGN   & 0.715  & Absorption system at 0.561,\citetalias{f2m07} & \\
040 & 10:49:18.24 & $+$15:44:58.09 &   5.33 & 26.66 & 23.48 & 20.71 & 19.41 & 17.90 & 15.45 & 5.26 & 2.45 & LRIS   & Galaxy  & 0.677  & \citetalias{f2m07} & \\
041 & 10:58:50.78 & $+$17:42:05.61 &   2.48 & 23.05 & 23.16 & 20.81 & 19.73 & 18.28 & 15.46 & 5.35 & 2.82 & none   & -       & -      & & \\
042 & 11:02:49.85 & $+$52:50:12.59 &  24.18 & 20.39 & 20.08 & 20.02 & 19.92 & 15.97 & 14.20 & 5.82 & 1.77 & SDSS   & QSO     & 0.690  & blue quasar & Sc05,VV06 \\
043 & 11:11:34.52 & $+$34:59:59.67 & 132.31 & 20.37 & 19.91 & 19.44 & 19.09 & 15.53 & 14.11 & 5.33 & 1.42 & none   & -       & -      & inverted radio spectrum & Ta84,+2 \\  
044 & 11:13:54.67 & $+$12:44:38.90 &   2.99 & 21.42 & 20.90 & 19.73 & 18.73 & 16.14 & 13.67 & 6.06 & 2.47 & ESI    & QSO     & 0.681  & HST image,\citetalias{f2m07} & VV06 \\
045 & 11:14:38.91 & $+$32:41:33.29 & 106.07 & 21.26 & 18.92 & 17.13 & 16.10 & 14.08 & 11.60 & 5.53 & 2.48 & Liter. & QSO     & 0.189  & IRAS source, ULIRG & He87,Kim98,+25 \\
046 & 11:18:11.06 & $-$00:33:41.87 &   1.30 & 23.04 & 21.41 & 20.46 & 19.52 & 17.06 & 14.58 & 5.88 & 2.48 & ESI    & QSO     & 0.686  & HST imaging,\citetalias{f2m07} & VV06 \\
047 & 11:21:48.09 & $+$21:32:41.35 &   5.81 & 21.34 & 21.19 & 20.95 & 20.48 & 17.40 & 15.40 & 5.55 & 2.00 & ESI    & QSO     & 0.834  & not very red & \\
048 & 11:39:26.18 & $+$27:43:56.64 &   2.63 & 21.17 & 20.39 & 19.40 & 18.90 & 17.05 & 13.04 & 6.36 & 4.01 & ESI    & NLAGN   & 0.394  & \citetalias{f2m07} & \\
049 & 11:49:08.91 & $+$28:24:35.03 & 241.11 & 22.09 & 21.38 & 20.93 & 20.44 & 17.03 & 15.43 & 5.50 & 1.60 & none   & -       & -      & flat spectrum radio source & Wi98,Sn02 \\
050 & 11:51:24.06 & $+$53:59:57.40 &   3.52 & 21.82 & 21.27 & 20.88 & 20.36 & 17.16 & 15.11 & 5.77 & 2.05 & ESI    & QSO     & 0.780  & HST image, Rosat source,\citetalias{f2m07} & VV06 \\
051 & 11:52:22.18 & $+$15:31:48.18 &  21.45 & 21.72 & 20.98 & 20.49 & 20.04 & 16.96 & 15.30 & 5.19 & 1.66 & SDSS   & Blazar  & 1.054? & featureless spectrum & \\
052 & 11:59:31.84 & $+$29:14:43.95 &1855.80 & 18.75 & 18.21 & 18.08 & 17.95 & 13.18 & 11.47 & 6.61 & 1.71 & Liter. & Blazar  & 0.729  & \citetalias{f2m07} & Wi83,Br84,+416 \\
053 & 12:01:11.14 & $-$03:32:19.64 &  55.27 & 22.61 & 20.99 & 19.91 & 19.10 & 16.68 & 14.87 & 5.04 & 1.81 & ESI    & Galaxy  & 0.755  & NLAGN?,\citetalias{f2m07} & \\
054 & 12:09:21.16 & $-$01:07:17.00 &   1.42 & 20.85 & 19.84 & 18.71 & 18.13 & 15.90 & 13.69 & 5.02 & 2.21 & Spex   & QSO     & 0.363  & only Near-IR spectrum,\citetalias{f2m07} & \\
\enddata
\end{deluxetable}

\clearpage

\setcounter{table}{2}
\begin{deluxetable}{lllrrccccccccccllcll}
\tabletypesize\tiny
\tablecaption{cont.}
\tablehead{
\colhead{No} & \colhead{R.A.} & \colhead{Dec} & \colhead{$F_{pk}$} & 
\colhead{u'} & \colhead{g'} &\colhead{r'} & \colhead{i'} &  \colhead{J} & 
\colhead{K} & \colhead{r'$-$K} & \colhead{J$-$K} & \colhead{Spect.} & 
\colhead{Type} & \colhead{$z$} & \colhead{comments} & \colhead{Reference}}
\startdata  
055 & 12:24:00.83 & $+$22:36:14.98 &   3.58 & 23.35 & 20.81 & 19.28 & 18.60 & 16.82 & 14.25 & 5.03 & 2.57 & none   & -       & -      & & \\
056 & 12:43:23.73 & $+$10:23:44.38 &   2.37 & 22.86 & 21.52 & 20.51 & 19.52 & 17.19 & 15.40 & 5.11 & 1.79 & ESI    & QSO     & 0.680  & composite Starburst & \\
057 & 12:48:47.16 & $+$05:31:30.79 &  10.62 & 21.43 & 21.09 & 20.69 & 20.34 & 18.08 & 14.63 & 6.06 & 3.45 & ESI    & QSO     & 0.740  & not very red & \\
058 & 12:50:57.75 & $+$64:06:32.73 &   1.61 & 23.75 & 22.27 & 20.86 & 19.25 & 16.83 & 15.25 & 5.61 & 1.58 & ESI    & Star    & 0.000  & & \\
059 & 12:56:15.29 & $-$02:00:23.58 &   2.26 & 22.85 & 23.13 & 21.18 & 19.86 & 17.00 & 15.13 & 6.05 & 1.87 & ESI    & QSO     & 0.835  & & VV06 \\
060 & 13:07:00.62 & $+$23:38:05.31 &   2.99 & 22.01 & 20.57 & 19.38 & 19.08 & 16.79 & 13.47 & 5.91 & 3.32 & ESI    & QSO     & 0.275  & \citetalias{f2m07} & Sm02,Ma03,+7 \\
061 & 13:09:17.03 & $+$60:42:08.86 &   4.24 & 22.79 & 22.72 & 21.57 & 20.53 & 17.89 & 14.80 & 6.77 & 3.09 & ESI    & QSO     & 0.641  & & \\
062 & 13:13:27.44 & $+$14:53:38.80 &   2.30 & 21.94 & 21.38 & 20.13 & 18.95 & 16.78 & 14.48 & 5.65 & 2.30 & SDSS   & QSO     & 0.584  & & \\
063 & 13:24:19.90 & $+$05:37:05.02 &   4.49 & 18.71 & 18.25 & 17.53 & 17.01 & 15.75 & 12.07 & 5.46 & 3.68 & SDSS   & QSO     & 0.203  & blue quasar, IRAS source & Lo88,Kim98,+55 \\
064 & 13:37:34.87 & $+$51:09:37.19 &   1.21 & 24.59 & 21.60 & 20.00 & 18.36 & 16.20 & 14.89 & 5.11 & 1.31 & SDSS   & Star    & 0.000  & & \\
065 & 13:40:39.66 & $+$05:14:19.39 &   8.89 & 20.84 & 19.93 & 19.06 & 18.53 & 16.16 & 13.26 & 5.80 & 2.90 & ESI    & NLAGN   & 0.264  & & \\
066 & 13:41:08.11 & $+$33:01:10.23 &  69.41 & 22.69 & 22.52 & 20.95 & 19.67 & 16.89 & 14.91 & 6.04 & 1.98 & ESI    & QSO     & 1.720  & BAL,\citetalias{f2m07} & Al86,Od95,VV06 \\
067 & 13:44:08.31 & $+$28:39:31.97 &   9.76 & 22.82 & 21.87 & 20.38 & 19.02 & 16.54 & 14.76 & 5.62 & 1.78 & SDSS   & QSO     & 1.770  & Mini-BAL, near M3,\citetalias{f2m07} & Od95,+2 \\
068 & 13:47:02.96 & $+$23:37:38.17 &  13.18 & 21.72 & 21.30 & 20.66 & 20.00 & 16.78 & 15.40 & 5.26 & 1.38 & none   & -       & -      & & \\
069 & 13:53:08.65 & $+$36:57:51.16 &   3.71 & 24.95 & 23.40 & 22.55 & 21.22 & 17.42 & 14.28 & 8.27 & 3.14 & ESI    & QSO     & 1.311  & \citetalias{f2m07} & VV06 \\
070 & 13:54:19.02 & $+$00:11:53.02 &   1.63 & 21.34 & 20.12 & 18.93 & 18.58 & 16.50 & 13.59 & 5.34 & 2.91 & ESI    & NLAGN   & 0.259  & & Li03 \\
071 & 13:57:39.30 & $+$35:19:05.16 &   1.83 & 21.84 & 21.57 & 20.84 & 20.01 & 17.26 & 15.68 & 5.16 & 1.58 & none   & -       & -      & & \\
072 & 14:15:22.83 & $+$33:33:06.52 &   5.21 & 21.50 & 20.59 & 19.31 & 18.62 & 16.73 & 14.29 & 5.02 & 2.44 & Spex   & QSO     & 0.416  & only Near-IR spectrum,\citetalias{f2m07} & \\
073 & 14:15:48.81 & $+$09:03:54.76 &  32.05 & 21.18 & 20.45 & 20.35 & 19.86 & 16.95 & 15.23 & 5.12 & 1.72 & ESI    & QSO     & 0.654  & Rosat & Br95,VV06,+2 \\
074 & 14:34:04.62 & $+$09:35:28.90 &   1.53 & 22.43 & 23.15 & 21.09 & 19.63 & 17.00 & 14.90 & 6.19 & 2.10 & ESI    & QSO     & 0.770  & wrong reference? & Mo07 \\
075 & 14:35:21.93 & $+$20:21:17.46 & 100.46 & 19.54 & 19.02 & 18.65 & 18.31 & 15.17 & 13.63 & 5.02 & 1.54 & Liter. & Blazar? & ?      & featureless spectrum & Ho96,Wi98,Sn02 \\
076 & 14:35:48.13 & $+$63:10:56.97 &  12.79 & 20.96 & 20.19 & 19.83 & 19.25 & 18.25 & 14.69 & 5.14 & 3.56 & KAST   & NLAGN   & 0.633  & & \\
077 & 14:39:39.50 & $+$15:55:55.16 &   6.34 & 24.87 & 22.54 & 21.31 & 20.17 & 17.29 & 15.41 & 5.90 & 1.88 & none   & -       & -      & & \\
078 & 14:50:57.21 & $+$53:00:08.03 &   1.67 & 22.98 & 21.10 & 20.63 & 19.55 & 16.66 & 15.10 & 5.53 & 1.56 & KAST   & Star    & 0.000  & part of FR2 jet? & \\
079 & 14:56:03.09 & $+$01:14:45.71 &   8.56 & 25.02 & 22.27 & 20.52 & 19.56 & 17.08 & 15.06 & 5.46 & 2.02 & SDSS   & QSO     & 2.378  & BAL & Ha02,Sc03,+2 \\  
080 & 14:57:38.13 & $+$07:49:54.70 & 195.15 & 21.65 & 21.07 & 20.48 & 19.95 & 16.92 & 15.42 & 5.06 & 1.50 & Liter. & Blazar  & 1.838  & High Frequency Peaker & Da00,Sn02,+8 \\
081 & 14:58:43.41 & $+$35:42:57.49 & 688.50 & 23.27 & 21.94 & 21.21 & 20.57 & 16.84 & 14.96 & 6.25 & 1.88 & ESI    & Galaxy  & 1.120  & Starburst,\citetalias{f2m07} & Ma82,Ma98,+5 \\
082 & 14:58:44.82 & $+$37:20:21.67 & 266.39 & 20.67 & 19.98 & 19.36 & 18.94 & 16.14 & 14.29 & 5.07 & 1.85 & SDSS   & Blazar? & 0.333? & High Frequency Peaker,\citetalias{f2m07} & Ho96,Da00,+16 \\
083 & 15:00:18.07 & $+$02:31:19.24 &   7.64 & 22.70 & 22.02 & 20.59 & 19.52 & 17.18 & 15.44 & 5.15 & 1.74 & SDSS   & QSO     & 1.501  & BAL & Sc03,VV06,+2 \\
084 & 15:04:16.51 & $+$06:01:43.75 &   3.39 & 22.76 & 21.81 & 20.30 & 18.89 & 16.60 & 15.27 & 5.03 & 1.33 & none   & -       & -      & & \\
085 & 15:06:53.05 & $+$42:39:23.19 & 297.59 & 21.24 & 20.60 & 20.03 & 19.23 & 16.79 & 14.99 & 5.04 & 1.80 & KAST   & Blazar? & 0.587? & Flat spectrum source & Sn02,VV06,+12 \\
086 & 15:07:18.10 & $+$31:29:42.33 &   7.79 & 22.32 & 21.77 & 20.66 & 19.84 & 16.79 & 15.16 & 5.50 & 1.63 & ESI    & QSO     & 0.988  & HST imaging & VV06 \\
087 & 15:11:45.67 & $+$12:59:06.40 &   1.49 & 26.19 & 23.11 & 21.81 & 20.53 & 17.59 & 15.48 & 6.33 & 2.11 & ESI    & Galaxy  & 0.664  & Starburst & \\
088 & 15:17:27.93 & $+$28:51:57.35 &   1.50 & 22.15 & 21.29 & 20.06 & 19.00 & 16.48 & 14.44 & 5.62 & 2.04 & Spex   & QSO     & 0.705  & only Near-IR spectrum & \\
089 & 15:22:27.82 & $-$01:51:22.03 &   1.72 & 22.96 & 21.60 & 19.76 & 18.50 & 16.45 & 14.70 & 5.06 & 1.75 & SDSS   & Star    & 0.000  & First source actually a jet? & \\
090 & 15:23:40.36 & $+$00:30:08.42 &   4.66 & 21.43 & 21.00 & 20.44 & 19.95 & 17.44 & 15.14 & 5.30 & 2.30 & ESI    & QSO     & 0.560  & & \\
091 & 15:25:54.81 & $+$53:04:01.70 &   2.96 & 22.43 & 21.43 & 20.04 & 19.38 & 17.08 & 15.03 & 5.01 & 2.05 & none   & -       & -      & & \\
092 & 15:40:43.74 & $+$49:23:23.89 &  32.99 & 23.34 & 21.53 & 20.49 & 19.40 & 17.46 & 15.18 & 5.31 & 2.28 & ESI    & QSO     & 0.696  & & \\
093 & 15:45:08.53 & $+$47:51:54.68 & 660.80 & 22.45 & 22.69 & 21.70 & 21.64 & 18.18 & 15.40 & 6.30 & 2.78 & Liter. & Blazar  & 1.277  & Superluminal flaring source & He95,Sn02,+19 \\
094 & 15:45:24.57 & $+$44:38:23.78 &   3.05 & 26.02 & 23.43 & 21.07 & 19.86 & 18.65 & 15.57 & 5.50 & 3.08 & ESI    & Galaxy  & 0.595  & & \\
095 & 15:48:42.02 & $+$09:13:32.92 &   2.71 & 23.38 & 22.37 & 20.84 & 19.67 & 17.50 & 15.77 & 5.07 & 1.73 & ESI    & QSO     & 1.524  & & VV06 \\
096 & 15:49:19.72 & $+$03:06:19.76 &   5.49 & 22.22 & 21.59 & 20.56 & 19.72 & 17.10 & 15.53 & 5.03 & 1.57 & ESI    & QSO     & 0.679  & & VV06 \\
097 & 15:51:02.81 & $+$08:44:01.54 &   3.14 & 25.26 & 22.50 & 20.66 & 19.55 & 16.71 & 14.14 & 6.52 & 2.57 & ESI    & QSO     & 2.505  & BAL & \\
098 & 15:52:37.39 & $+$05:43:43.21 &   4.48 & 23.99 & 22.39 & 20.78 & 19.34 & 17.00 & 15.29 & 5.49 & 1.71 & ESI    & Star    & 0.000  & & \\
099 & 15:54:54.87 & $+$23:31:51.02 &   2.59 & 22.95 & 22.44 & 20.83 & 19.55 & 17.00 & 15.15 & 5.68 & 1.85 & none   & -       & -      & & \\
100 & 15:57:08.31 & $+$03:06:28.94 &   1.35 & 24.71 & 22.08 & 19.99 & 19.03 & 16.62 & 14.92 & 5.07 & 1.70 & ESI    & Galaxy  & 0.463  & & \\
101 & 16:00:01.07 & $+$28:29:29.26 &   1.20 & 23.95 & 23.62 & 22.59 & 20.32 & 17.23 & 15.22 & 7.37 & 2.01 & ESI    & Star    & 0.000  & Very faint & \\
102 & 16:12:08.47 & $+$10:48:37.30 &   2.34 & 23.55 & 22.11 & 20.47 & 19.72 & 17.04 & 15.40 & 5.07 & 1.64 & none   & -       & -      & & \\
103 & 16:12:58.24 & $+$38:50:13.17 &   1.07 & 22.70 & 21.85 & 20.56 & 19.58 & 17.46 & 15.34 & 5.22 & 2.12 & ESI    & Galaxy  & 0.655  & Starburst, perhaps composite & \\
104 & 16:15:47.90 & $+$03:18:50.83 &  17.62 & 21.85 & 20.59 & 19.92 & 19.38 & 16.53 & 14.87 & 5.05 & 1.66 & ESI    & QSO     & 0.424  & radio source has jets & \\
105 & 16:18:09.72 & $+$35:02:08.53 &  13.73 & 20.66 & 20.00 & 19.07 & 18.30 & 16.76 & 14.06 & 5.01 & 2.70 & KAST   & NLAGN   & 0.446  & 2MASS red AGN?,\citetalias{f2m07} & Hu06 \\  
106 & 16:24:48.55 & $+$29:19:16.78 &   4.43 & 22.41 & 21.88 & 20.19 & 19.01 & 16.82 & 14.62 & 5.57 & 2.20 & ESI    & Galaxy  & 0.536  & & \\
107 & 16:35:47.04 & $+$11:53:12.26 &   1.78 & 23.35 & 22.23 & 20.82 & 19.74 & 16.53 & 15.17 & 5.65 & 1.36 & none   & -       & -      & & \\
108 & 16:38:42.99 & $+$27:07:05.34 &   4.17 & 23.95 & 22.73 & 21.68 & 20.37 & 17.16 & 15.22 & 6.46 & 1.94 & ESI    & QSO     & 1.692  & Mini-BAL & \\
\enddata
\end{deluxetable}

\clearpage

\setcounter{table}{2}
\begin{deluxetable}{lllrrccccccccccllcll}
\tabletypesize\tiny
\tablecaption{cont.}
\tablehead{
\colhead{No} & \colhead{R.A.} & \colhead{Dec} & \colhead{$F_{pk}$} & 
\colhead{u'} & \colhead{g'} &\colhead{r'} & \colhead{i'} &  \colhead{J} & 
\colhead{K} & \colhead{r'$-$K} & \colhead{J$-$K} & \colhead{Spect.} & 
\colhead{Type} & \colhead{$z$} & \colhead{comments} & \colhead{Reference}}
\startdata  
109 & 16:39:13.02 & $+$53:28:49.80 &  12.31 & 21.55 & 21.12 & 20.05 & 19.30 & 17.07 & 15.04 & 5.01 & 2.03 & none   & -       & -      & in cluster A2220 z=0.11 & Ha80 \\
110 & 16:50:37.43 & $+$32:42:19.01 &   8.24 & 22.31 & 22.27 & 20.70 & 19.89 & 17.15 & 15.38 & 5.32 & 1.77 & ESI    & QSO     & 1.059  & center of FR2 & Ma78 \\
111 & 16:56:47.11 & $+$38:21:36.72 &   4.12 & 22.99 & 23.44 & 22.03 & 20.65 & 17.38 & 15.12 & 6.91 & 2.26 & ESI    & QSO     & 0.732  & NLAGN in optical, HST imaging,\citetalias{f2m07} & VV06 \\
112 & 17:08:02.53 & $+$22:27:25.70 &   1.31 & 22.78 & 21.82 & 20.22 & 19.65 & 16.96 & 14.59 & 5.63 & 2.37 & ESI    & QSO     & 0.377  & \citetalias{f2m07} & \\
113 & 17:15:59.79 & $+$28:07:16.68 &   1.59 & 22.01 & 21.26 & 20.47 & 19.27 & 17.15 & 14.63 & 5.84 & 2.52 & SDSS   & NLAGN   & 0.523  & broad 2MASS red AGN? & Sm02,Ma03,+7 \\
114 & 17:20:27.58 & $+$61:56:57.63 &   2.85 & 26.30 & 23.35 & 21.64 & 20.22 & 17.15 & 15.20 & 6.44 & 1.95 & ESI    & QSO     & 0.727  & & \\
115 & 21:13:01.49 & $-$06:35:25.73 &   5.26 & 23.06 & 22.54 & 20.89 & 19.80 & 17.44 & 15.11 & 5.78 & 2.33 & ESI    & NLAGN   & 0.481  & & \\
116 & 21:52:14.08 & $-$00:28:32.59 &   1.15 & 22.10 & 22.00 & 20.90 & 20.05 & 16.94 & 15.56 & 5.34 & 1.38 & SDSS   & QSO     & 0.867  & & VV06 \\
117 & 21:59:28.05 & $+$01:10:25.00 &  11.94 & 25.28 & 22.55 & 20.41 & 19.14 & 16.95 & 15.23 & 5.18 & 1.72 & SDSS   & Galaxy  & 0.635  & & \\
118 & 22:24:38.35 & $-$00:07:50.95 &   1.41 & 24.50 & 21.95 & 20.09 & 19.32 & 16.87 & 14.97 & 5.12 & 1.90 & ESI    & Galaxy  & 0.452  & \citetalias{f2m07} & \\
119 & 22:55:23.33 & $+$00:49:43.07 &  13.55 & 25.17 & 24.86 & 23.24 & 21.85 & 17.16 & 15.33 & 7.91 & 1.83 & none   & -       & -      & & \\
120 & 23:35:09.42 & $-$10:22:44.04 &   6.55 & 23.26 & 21.77 & 20.96 & 20.00 & 17.30 & 15.63 & 5.33 & 1.67 & none   & -       & -      & FR2 & \\
121 & 23:39:03.84 & $-$09:12:20.99 &   4.34 & 23.05 & 21.02 & 19.28 & 18.17 & 15.75 & 14.12 & 5.16 & 1.63 & SDSS   & QSO     & 0.660  & \citetalias{f2m07} & Be01,Sc05,+2 \\
122 & 23:55:28.34 & $-$09:23:13.38 &   3.67 & 22.77 & 22.16 & 20.60 & 19.66 & 17.22 & 15.33 & 5.27 & 1.89 & Spex   & QSO     & 0.512  & only Near-IR spectrum & \\
\enddata

\tablecomments{Units of right ascension are hours, minutes, and seconds, and 
units of declination are degrees, arcminutes, and arcseconds. We use FIRST 
coordinates, which are in J2000 epoch.}

\tablerefs{(Al86) \cite{al86}; (Am88) \cite{am88}; (Ba06) \cite{ba06}; 
(Be96) \cite{be96}; (Be01) \cite{be01}; (Br84) \cite{br84}; 
(Br95) \cite{br95}; (Da00) \cite{da00}; (deG92) \cite{deg92};
(Gr02) \cite{gr02}; (Ha80) \cite{ha80}; (Ha02) \cite{ha02}; 
(He87) \cite{he87}; (He95) \cite{he95}; (Ho96) \cite{ho96}; 
(Hu03) \cite{hu03}; (Hu06) \cite{hu06}; (Kh04) \cite{kh04}; 
(Kim98) \cite{kim98}; (Ki99) \cite{ki99}; (La02) \cite{la02}; 
(Le01) \cite{le01}; (Li03) \cite{li03}; (Lo88) \cite{lo88}; 
(Ma78) \cite{ma78}; (Ma82) \cite{ma82}; (Ma98) \cite{ma98}; 
(Ma03) \cite{ma03}; (Mo07) \cite{mo07}; (Od95) \cite{od95}; 
(Ri03) \cite{ri03}; (Sc03) \cite{sc03}; (Sc05) \cite{sc05}; 
(Sm02) \cite{sm02}; (Sn01) \cite{sn01}; (Sn02) \cite{sn02}; 
(Ta84) \cite{ta84}; (Ur05) \cite{ur05}; (VV06) \cite{vv06}; 
(Wi83) \cite{wi83}; (Wi98) \cite{wi98}; (Wi02) \cite{wi02}; 
(Za03) \cite{za03}}

\end{deluxetable}

\clearpage

\end{landscape}

\end{document}